\documentclass[prb,twocolumn,aps,superscriptaddress,10pt]{revtex4-2}
\usepackage{graphicx}
\usepackage{dcolumn}
\usepackage{amsmath}
\usepackage{amssymb}
\usepackage{bm}
\usepackage{mathtools}
\usepackage{physics}
\usepackage[utf8]{inputenc}
\usepackage[usenames, dvipsnames]{color}
\usepackage[english]{babel}
\usepackage[T1]{fontenc}
\usepackage{bbm}
\usepackage{float}
\usepackage{verbatim}
\usepackage[caption=false]{subfig}
\usepackage{xfrac}
\usepackage{upgreek}
\usepackage{makecell}
\usepackage{comment}
\usepackage{placeins}
\usepackage[normalem]{ulem}
\usepackage{hyperref}
\usepackage{xcolor}
\hypersetup{colorlinks = true, linkcolor=blue, citecolor=blue, urlcolor=blue}

  \makeatletter
    \renewcommand\@make@capt@title[2]{%
     \@ifx@empty\float@link{\@firstofone}{\expandafter\href\expandafter{\float@link}}%
      {\textsc{#1}}\@caption@fignum@sep#2\quad}%
    \makeatother
\begin{document}
\title{Do single-shot projective readouts necessarily estimate the $T_1$ lifetime ?}

\author{Aparajita Modak}
\affiliation{Department of Electrical Engineering, Indian Institute of Technology Bombay, Powai, Mumbai-400076, India}

\author{Sundeep Kapila}
\affiliation{Department of Electrical Engineering, Indian Institute of Technology Bombay, Powai, Mumbai-400076, India}
\author{Bent Weber}
\affiliation{Division of Physics and Applied Physics, School of Physical and Mathematical Sciences,
Nanyang Technological University, Singapore 637371, Singapore}
\author{Klaus Ensslin}
\affiliation{Laboratory for Solid State Physics, ETH Zurich, CH-8093 Zurich, Switzerland}
\author{Guido Burkard}
\affiliation{Department of Physics, University of Konstanz, D-78457, Germany}
\author{Bhaskaran Muralidharan}
\email{Corresponding author: bm@ee.iitb.ac.in}
\affiliation{Department of Electrical Engineering, Indian Institute of Technology Bombay, Powai, Mumbai-400076, India} 
\date{\today}
\begin{abstract}
When single-shot qubit readout protocols are adapted for multilevel systems, theoretical $T_1$ lifetime calculations often fall short of capturing the experimental lifetime trends. We identify {\it extrinsic} population dynamics as the fundamental origin of this disparity, establishing that the lifetime estimates can, in certain operating regions, be distinct from the intrinsic $T_1$ time. We clarify these aspects with an integrated theory to address recent measurements {\bf{[Nat. Nano, 20, 494, (2025)]}} on spin-valley states in bilayer graphene. While confirming that phonon and Johnson noise are the dominant intrinsic sources, we show that the inclusion of extrinsic factors provide the critical match to the experimental estimates. The extrinsic factors also effectuate violations of generalized Mathiessen's rules. With an improved handle on the design space, a revised readout protocol to estimate the $T_1$ lifetime of the valley qubit is proposed.
\end{abstract}
\maketitle
\begin{figure*}[thpb]
\centering
\includegraphics[width=\textwidth,height=0.45\textwidth,, keepaspectratio]{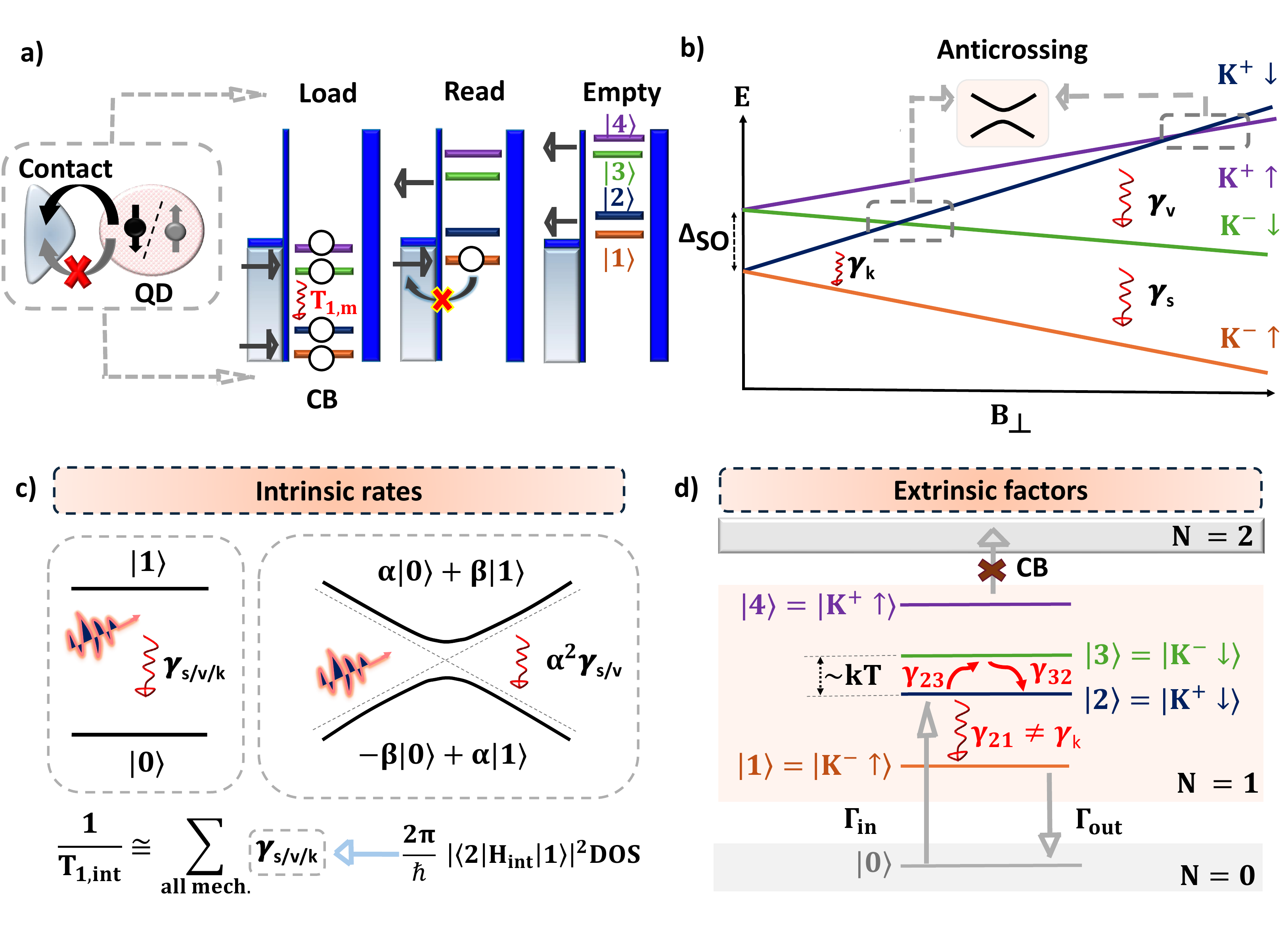}
\caption{ Preliminaries. (a) ERT based on energy-selective tunneling between the dot and the reservoir. 
(b) Representative energy spectrum of the four spin--valley states as a function of $B_{\perp}$. The first anticrossing occurs when $E_Z$ equals $\Delta_{\mathrm{SO}}$, with $t_v$ inducing hybridized eigenstates. A second, weaker anticrossing appears at higher fields due to intravalley mixing, leading to a modified state composition. The intrinsic relaxation channels, spin ($\gamma_s$), valley ($\gamma_v$), and Kramers ($\gamma_k$) are indicated. (c) Intrinsic pathway away from, and near the anticrossing, which involves state--mixing. (d) Extrinsic factors: population dynamics which modify the experimentally inferred lifetime.} 
\label{fig:Fig1}
\end{figure*}
\begin{figure*}[t]
\centering
\includegraphics[width=0.7\textwidth,height=0.45\textwidth,keepaspectratio]{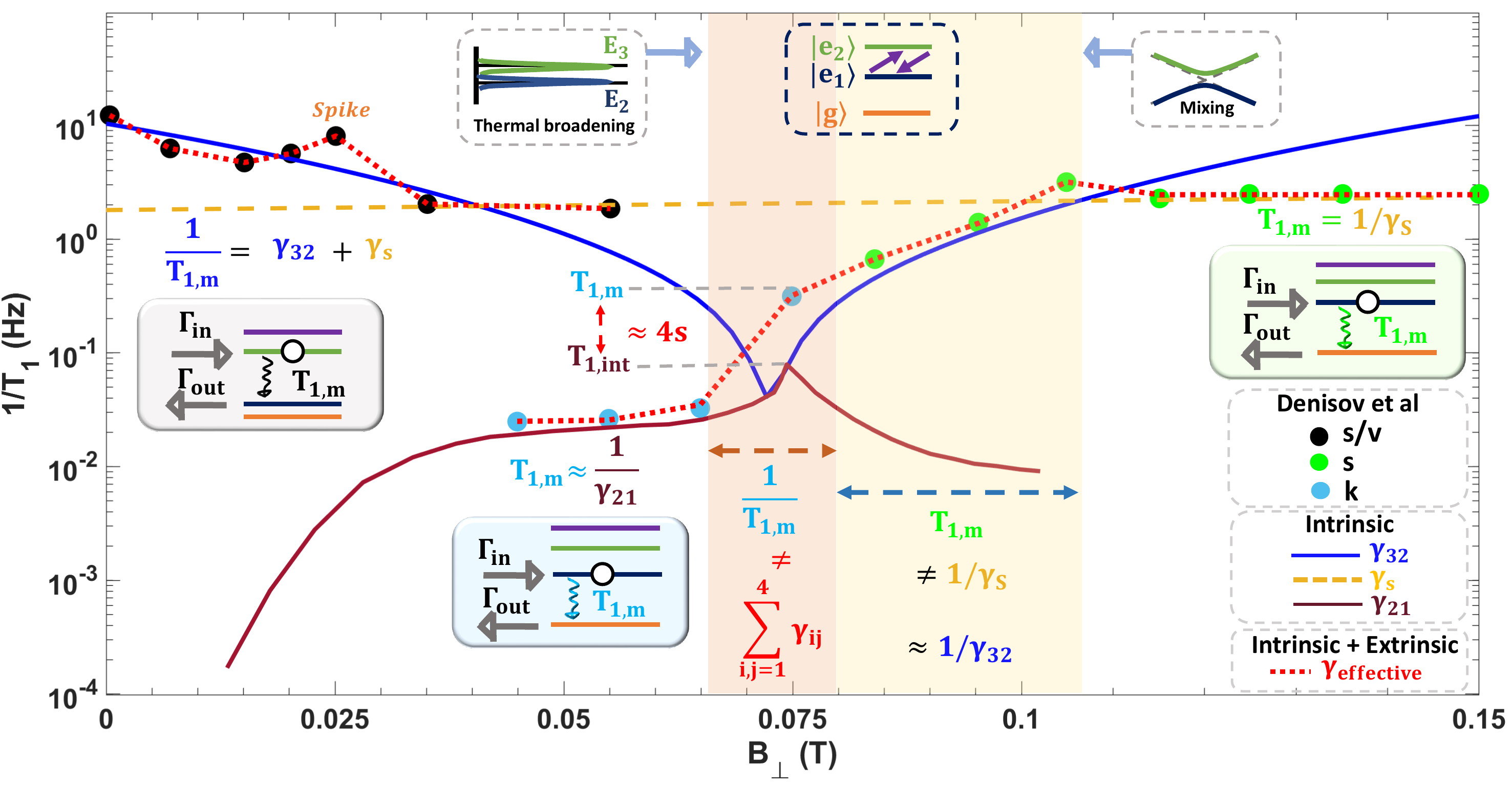}
\caption{Experimental match, intrinsic and extrinsic factors. The dependence of intrinsic spin, valley and Kramers lifetimes on $B_\perp$ with contributions from phonon and Johnson noise in the presence of $t_v$. The experimentally measured lifetimes ($T_{1m}$) and the corresponding readout protocols (inset). Shaded regions mark the regions where $t_v$ and thermal broadening become relevant. The effective lifetimes ($\gamma_{\mathrm{effective}}^{-1}$) obtained from the integrated framework capture the observed behavior. Parameters used are $g_1=50~\mathrm{eV}$ and $g_2=3~\mathrm{eV}$ (s, k), $g_1=50~\mathrm{eV}$ and $g_2=2.8~\mathrm{eV}$ (v), $J_s=1.8~\mathrm{Hz}$, $J_v=2.5~\mathrm{Hz}$, $J_k=0.02~\mathrm{Hz}$, $R=30~\mathrm{nm}$, $U_0=40~\mathrm{meV}$, $\Delta_{\mathrm{SO}}=64~\mu\mathrm{eV}$, $t_v=1~\mu\mathrm{eV}$, and $g_v=14.5$~\cite{denisov2025spin}.}
\label{fig:Fig2}
\end{figure*}

\indent  In any physical realization of quantum computation \cite{IBMQuantum}, the $T_1$ lifetime of a qubit is a critical performance parameter as it asserts the immunity to environmental noise. It is defined as the longitudinal relaxation time of a quantum two-state system (QTS, qubit)  as a result of {\it{intrinsic}} ambient noise \cite{Blum:2012:DMTA,IBMQuantum,Qiskit}. In the Markovian case, it limits the dephasing time $T_2$, with $T_2\leqslant2T_1$ \cite{nielsenchuang10th}. \\
\indent In semiconductor quantum dots \cite{RevModPhys.79.1217,zwanenburg2013silicon,RevModPhys.95.025003}, the $T_1$ time is estimated by employing the single-shot projective readout, also called the Elzerman readout technique \cite{elzerman2004single,morello2010single,hanson2007spins}. The implementation \cite{hanson2007spins} on a QTS such as a Zeeman spin qubit \cite{elzerman2004single} yields an accurate estimate of the spin-$T_1$ time. The presence of closely placed excited states (ES) near the probed states of interest, degeneracies and level anticrossings, however, are inevitable in semiconductor systems of current interest, such as Si, 2D-material platforms like bilayer graphene (BLG) and transition metal dichalcogenides \cite{denisov2025spin,banszerus2022spin,banszerus2025phonon,garreis2024long,PRXQuantum.3.020343,elzerman2004single,xiao2010measurement,simmons2011tunable,petit2018spin}, due to the coupling of spins and inequivalent valleys.  \\ 
\indent The ERT can be adapted for the multilevel case also \cite{yang2013spin} via energy selective {\it{readout protocols}} \cite{hanson2007spins} that probe the two states of interest as schematized in Fig.~\ref{fig:Fig1}(a). The load and read cycles are facilitated by single charge tunneling events involving a fermionic reservoir. For a viable $T_1$ estimate, the energy separation must hence be sufficiently larger than the thermal energy.   
Despite satisfying the aforesaid criterion, we establish here that the effective lifetimes obtained from a multilevel energy selective ERT \cite{yang2013spin,denisov2025spin} can be distinct from the $T_1$ time. \\
\indent Recent measurements \cite{denisov2025spin} on bilayer graphene quantum dots (BLG-QDs) provide the requisite backdrop while simultaneously facilitating a testbed for a necessary theory-experiment synergy. Figure~\ref{fig:Fig1}(b) shows a schematic of the energy landscape of a singly occupied BLG-QD with an applied out-of-plane magnetic field $B_{\perp}$. This platform can host the spin, valley and the spin-valley (Kramers) qubit, whose operating points can be tuned as $B_{\perp}$ is varied. Among these, the Kramers qubit promises ultra-long $T_1$ times \cite{denisov2025spin,banszerus2021spin}, owing to the need for a simultaneous flip of the spin and the valley state near zero magnetic fields; due to what is known as the van Vleck cancellation \cite{vanvleck_blg}.\\
\indent To distinguish the $T_1$ time from an effective lifetime, especially around degeneracies and anticrossings, one has to disentangle the intrinsic decay pathways from the {\it{extrinsic}} ones that include stochastic loading, hybridization, and the proximity to nearby ES. The intrinsic point of view that defines the $T_1$ time is schematized in Fig.~\ref{fig:Fig1}(c), while the extrinsic factors that alter the population dynamics are depicted via the state transition diagram in Fig.~\ref{fig:Fig1}(d). To faithfully reproduce the experimentally estimated lifetimes, as shown in Fig.~\ref{fig:Fig2}, we develop a two-tier approach, that blends in the extrinsic factors. We integrate the extrinsic dynamics described by the master equation (ME) formulation \cite{RevModPhys.79.1217,muralidharan2007generic,shandilya2025unified} over the intrinsic noise models evaluated using the Fermi's golden rule (FGR) approach \cite{khaetskii2001spin,golovach2004phonon,RevModPhys.95.025003,wang2024valley,struck2010effective,droth2013electron}.  \\
\indent Our intrinsic calculations indeed ratify the experimentally measured ultralong Kramers $T_1$ times and that the phonon and Johnson noise are the dominant sources of noise. However, Fig.~\ref{fig:Fig2} also points out that the intrinsic treatment alone underestimates the relaxation rate specifically in the $B_{\perp}$ range of the shaded regions. As an extreme case marked in Fig.~\ref{fig:Fig2}, the experiment reports a Kramers lifetime approximately $4~\mathrm{s}$ longer than the intrinsic $T_1$ at $B_{\perp} = 0.075~\mathrm{T}$, which is around the valley anti-crossing region.\\
\indent Cementing the role of both intrinsic and extrinsic contributions, we refer back to the theory–experiment agreement in Fig.~\ref{fig:Fig2}, where the experimentally measured qubit lifetimes ($T_{1,m}$) and the associated measurement protocols (inset) are shown.
The blue and green data points partially follow intrinsic trends, whereas the black data points follow a combination of $\gamma_s$ and $\gamma_v$, the valley and spin relaxation rates respectively. The shaded regions around the anticrossing require the inclusion of extrinsic analysis. The additional pathways activated by these contributions also lead to a violation of the generalized Matthiessen’s harmonic sum rule of the individual lifetimes.\\
\indent {\it{Preliminaries.}}  Neglecting the Rashba spin-orbit coupling \cite{abdelouahed2010spin,min2006intrinsic}, the effective Hamiltonian of the BLG-QD in a homogeneous $B_{\perp}$ is,
\begin{align}
\hat{H}_{QD} &= \epsilon_0 \hat{n}
+ t_v \sum_{\sigma}
\left(\hat{c}^\dagger_{K^{+}\sigma}\hat{c}_{K^{-}\sigma} + \text{h.c.}\right) \nonumber\\
&\quad + \frac{1}{2}\sum_{\tau,\sigma}
\hat{c}^\dagger_{\tau\sigma}
\left[\Delta_{\rm SO}\,\tau_z s_z
+ \mu_B B_\perp (g_s s_z + g_v \tau_z)\right]
\hat{c}_{\tau\sigma} \nonumber\\
&\quad + \frac{U}{2}\left(\hat{n}^2 - \hat{n}\right).
\end{align}
where $\epsilon_0$ is the on-site energy, $\hat{c}^\dagger_{\tau\sigma} (\hat{c}_{\tau \sigma})$ is the creation (annihilation) operator of an electron with spin $\sigma$ residing in a valley $\tau =K^{\pm}$. The term $U$ is the single-electron charging energy, $\hat{n}$ is the total number operator, $\Delta_{\mathrm{SO}}$, the intrinsic Ising SOC, $g_{s(v)}$, the spin (valley) $g$-factor and $t_v$ is the intervalley mixing parameter. The Ising SOC term $\Delta_{\mathrm{SO}} \tau_z s_z$ preserves both spin and valley quantum numbers, while the valley Zeeman coupling $\propto g_v B_{\perp} \tau_z$ shifts $K^+$ and $K^-$ states oppositely. We now define the generalized Zeeman energy $E_Z= \mu_B B_\perp (g_s s_z + g_v \tau_z)$, which combines the spin and valley Zeeman effects.  \\
\indent To make the role of state mixing (hybridization) explicit, we begin with the spin–valley product states, $\lvert 1\rangle=\lvert K^{-}\uparrow \rangle,\lvert 2\rangle=\lvert K^{+}\downarrow \rangle,\lvert 3\rangle=\lvert K^{-}\downarrow \rangle, \lvert 4\rangle=\lvert K^{+}\uparrow \rangle\}$. 
Intervalley mixing $t_v$ 
them as 
\begin{align}
\begin{aligned}
\lvert 1 \rangle &= \lvert K^{-} \uparrow \rangle, \\
\lvert \bar{2} \rangle &= \sin(\tfrac{\theta}{2})\, \lvert K^{+} \downarrow \rangle - \cos(\tfrac{\theta}{2})\, \lvert K^{-} \downarrow \rangle, \\
\lvert \bar{3} \rangle &= \cos(\tfrac{\theta}{2})\, \lvert K^{+} \downarrow \rangle + \sin(\tfrac{\theta}{2})\, \lvert K^{-} \downarrow \rangle, \\
\lvert 4 \rangle &= \lvert K^{+} \uparrow \rangle.
\end{aligned}
\end{align}
Here, $\theta$ is the mixing angle determined by the interplay of $\Delta_{\mathrm{SO}}$, the magnetic field-induced Zeeman energy $E_Z$, and $t_v$. Perturbation theory yields
$\cos^2\!\left(\frac{\theta}{2}\right)
=\frac{\sqrt{(\Delta_{\mathrm{SO}} - E_Z)^2 + 4t_v^2} - (\Delta_{\mathrm{SO}} - E_Z)}
{2\sqrt{(\Delta_{\mathrm{SO}} - E_Z)^2 + 4t_v^2}}$. At the anticrossing, $\Delta_{\mathrm{SO}} = E_Z$, and the level repulsion is determined by $t_v$.\\
\indent \textit{Intrinsic lifetimes.} 
The spin, valley, and Kramers qubits in BLG quantum dots are subject to several electronic noise sources, most notably, phonon noise, Johnson noise, and $1/f$ charge noise~\cite{huang2014electron,huang2021fast,hosseinkhani2021relaxation,tse2009energy}. At GHz frequencies of relevance here, phonon and Johnson noise dominate the relaxation dynamics. In a centrosymmetric crystal like BLG~\cite{viljas2010electron,borysenko2011electron}, the electron–phonon coupling (EPC) arises primarily from the deformation-potential (DP) and bond–length–change (BLC) mechanisms. We include contributions due to both longitudinal acoustic (LA) and transverse acoustic (TA) phonon branches. The in-plane EPC is described by the Hamiltonian ~\cite{ando2005theory,wang2024valley}
\begin{equation}
\hat{H}_{\mathrm{EPC}} = K
\begin{pmatrix}
g_1 a_1 & g_2 a_2^* \\
g_2 a_2 & g_1 a_1
\end{pmatrix}
\left(e^{i \mathbf{q}\cdot \mathbf{r}} \hat{b}^\dagger_{mq} - e^{-i \mathbf{q}\cdot \mathbf{r}} \hat{b}_{mq} \right),
\end{equation}
where $K=\tfrac{q}{\sqrt{A \rho \, \omega_{qm}}}$, $A$ is the area, $\rho$ is the mass density, $\omega_{\mathbf{q},m}$ is the phonon frequency for the wave vector $\mathbf{q}$ and the branch $m$, $g_{1(2)}$ is the coupling constant of DP(BLC), $a_{1(2)}$ are the sublattice amplitudes and $\hat{b}^\dagger_{mq}$ ($\hat{b}_{mq}$) is the phonon creation (annihilation) operator. \\
\indent In hBN-encapsulated BLG, coupling to out-of-plane flexural (ZA) phonon modes is strongly suppressed by mechanical clamping, leaving in-plane acoustic phonons as the dominant relaxation channel. Applying the FGR, the intrinsic phonon-assisted transition rates are obtained as follows (see Supplementary Note~I):
\begin{equation}
\gamma_{\alpha}(\omega)
=
\sum_{m=\mathrm{LA,TA}}
\gamma^{\mathrm{ph}}_{m\alpha}(\omega)
\left(\frac{\omega}{\omega_{\mathrm{SO}}}\right)^{n_{\alpha}}
+
J_{\alpha}\,\gamma^{J}_{\alpha}(\omega),
\end{equation}
where $\alpha \in \{s,v,k\}$, with $n_v = 4$, $n_k = 6$, and
$n_s = 2~(\mathrm{DP,\,LA})$ or $n_s = 4~(\mathrm{BLC,\,LA,\,TA})$. The spin–orbit splitting frequency is denoted by $\omega_{\mathrm{SO}}$. The matrix elements $\gamma^{\mathrm{ph}}_{\alpha m}(\omega)$ arise from in-plane acoustic modes, vary only weakly with $B_{\perp}$ at low fields and are set primarily by the dot size and sound velocities. In contrast, $\gamma^{\mathrm{J}}_{\alpha}(\omega)$ describes the relaxation driven by Johnson noise, and $J_{\mathrm{\alpha}}$ encodes the device-specific electromagnetic coupling parameters.\\
\indent  
For small $B_\perp$, 
$t_v$ is the dominant source of hybridization, and the transition rates for the four states cane be written as (see Supplementary Note II):
\begin{align}
\begin{aligned}
\gamma_{\bar{2}1} &= \gamma_{s} \cos^2(\tfrac{\theta}{2}) + \gamma_{k}\sin^2(\tfrac{\theta}{2}), \\
\gamma_{\bar{3}1} &= \gamma_{s} \sin^2(\tfrac{\theta}{2}) + \gamma_{k}\cos^2(\tfrac{\theta}{2}), \\
\gamma_{\bar{3}\bar{2}} &= \gamma_{v} \cos^2\theta, \\
\gamma_{4\bar{3}} &= \gamma_{s} \cos^2(\tfrac{\theta}{2}) + \gamma_{k}\sin^2(\tfrac{\theta}{2}), \\
\gamma_{4\bar{2}} &= \gamma_{s} \sin^2(\tfrac{\theta}{2}) + \gamma_{k}\cos^2(\tfrac{\theta}{2}), \\
\gamma_{41} &= \gamma_{v}.
\end{aligned}
\end{align}
Here, $\bar{2}$ and $\bar{3}$ denote the excited hybridized states near the anticrossing. The above illustrates that state-mixing modulates the relaxation rates $\gamma_\alpha$ near the anticrossing~\cite{huang2014spin,yang2013spin,hosseinkhani2022theory,khaetskii2001spin}. In Fig.~\ref{fig:Fig2}, we hence note that a finite $t_v$ produces a dip in $\gamma_{v}$, whereas $\gamma_{s}$ remains essentially constant in this regime, as the spin splitting $\Delta E_{s}$ varies only weakly with $B_{\perp}$. However, $\gamma_{v}$ exhibits a scaling of $B_{\perp}^{6}$. This trend is altered near the anticrossing, where $t_v$ modifies the relaxation matrix elements as noted in the above expressions. \\
\indent \textit{Extrinsic contributions.} The ME framework captures the nonequilibrium populations in the eigenstate basis. Coherences between eigenstates can be neglected because they decay much faster than the intrinsic relaxation processes driven by noise and charge tunneling. 
A quantum master-equation treatment is, however, required when a transverse electromagnetic field is applied~\cite{Brouwer,Timm,BM_Grifoni,campaioli2024quantum,ferguson2017long}.\\
\begin{figure}[h]
\centering
\includegraphics[width=0.82\columnwidth,height=\columnwidth, keepaspectratio]{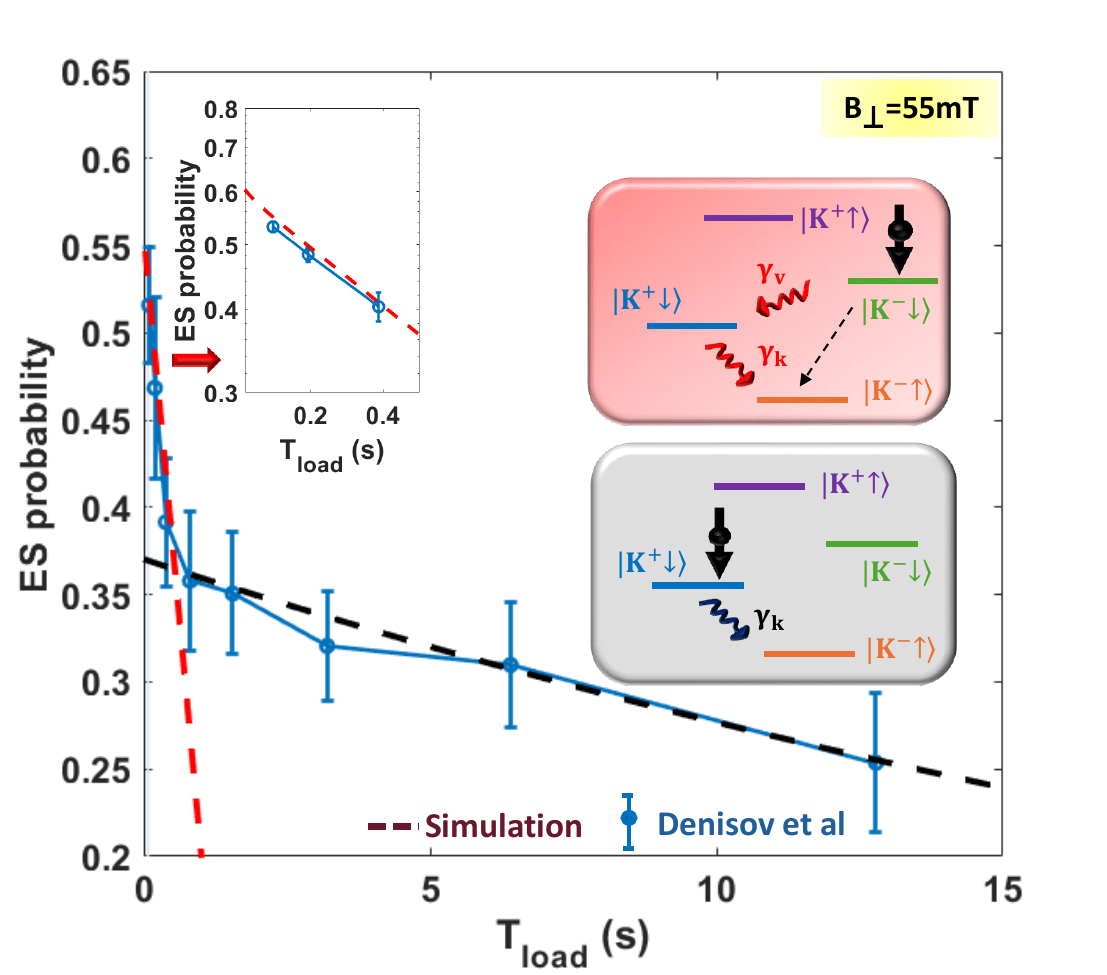}\caption{Extrinsic effects. Bi-exponential distribution of the ES probability versus $T_{\mathrm{load}}$ at $B_{\perp}=55~\mathrm{mT}$. Charge jumps occasionally load the electron into the excited state $\lvert K^{-}\!\downarrow\rangle$ (top), producing fast valley-mediated relaxation. In the absence of such jumps, the system relaxes through the Kramers channel $\lvert K^{+}\!\downarrow\rangle$ (bottom).}
\label{fig:Fig3}
\end{figure}
\indent To account for the load and read operations involving the contacts, the total Hamiltonian $\hat{H} = \hat{H}_{QD} + \hat{H}_C + \hat{H}_{T}$, where $\hat{H}_C$ and  $\hat{H}_{T}$ are the contact and tunneling Hamiltonian respectively, is considered. Here, $\hat{H}_{T}$ is treated perturbatively ~\cite{Brouwer,Timm,BM_Grifoni,campaioli2024quantum,ferguson2017long} up to second order to obtain the ME for the evolution of the diagonal elements (state probabilities) in the BLG-QD subspace. \\
\indent The BLG-QD subspace is described using a five state Fock space comprising four one-electron ($N=1$) states and one unoccupied ($N=0$) state. The tunnel coupling induces transitions between $N=0$ and $N=1$ space (Fig.~\ref{fig:Fig1}(d)), as a part of the load and read processes. Intrinsic transitions are described by $R_{ij}=\gamma_{ij}$, while tunneling to and from the contact is given by $R_{ij}=\Gamma_{ij}$. The occupations $P_i$ evolve according to 
\begin{equation}
\frac{dP_i}{dt} = - \sum_{j} R_{ij} P_i + \sum_{j} R_{ji} P_j,
\end{equation}
with $i,j=0,1,2(\bar{2}),3(\bar{3}),4$ and $\sum_i P_i=1$. Here, $i(j)=0$ labels the $N=0$ state. The dot–lead tunneling is accounted for using $\Gamma_i=\gamma_b f_i$, where $f_i=[1+\exp((E_i-\mu_F)/k_BT_e)]^{-1}$, $\gamma_b$ is the bare tunneling rate, $\mu_F$ is the electrochemical potential of the contact \cite{muralidharan2007generic}, and $T_e$ is the electron temperature. The transients of the readout processes are captured with respect to the load time, $T_{load}$ (See Supplementary Note IV). \\
\indent As a first exposition on extrinsic effects, Fig.~\ref{fig:Fig3} explains the experimentally observed bi-exponential decay of the ES population with $T_{\mathrm{load}}$. For the relevant parameters ($\gamma=15$~Hz, $B_\perp=55$~mT, and $T_e=30$~mK), the second ES lies well outside the thermal window ($\Delta \epsilon \gg 3k_{\mathrm{B}}T_e$) and cannot be accessed otherwise. In practical systems, stochastic charge fluctuations enable occasional loading into either ES with comparable probability~\cite{denisov2025spin}.\\
\begin{figure}[h]
\centering
\includegraphics[width=0.87\columnwidth]{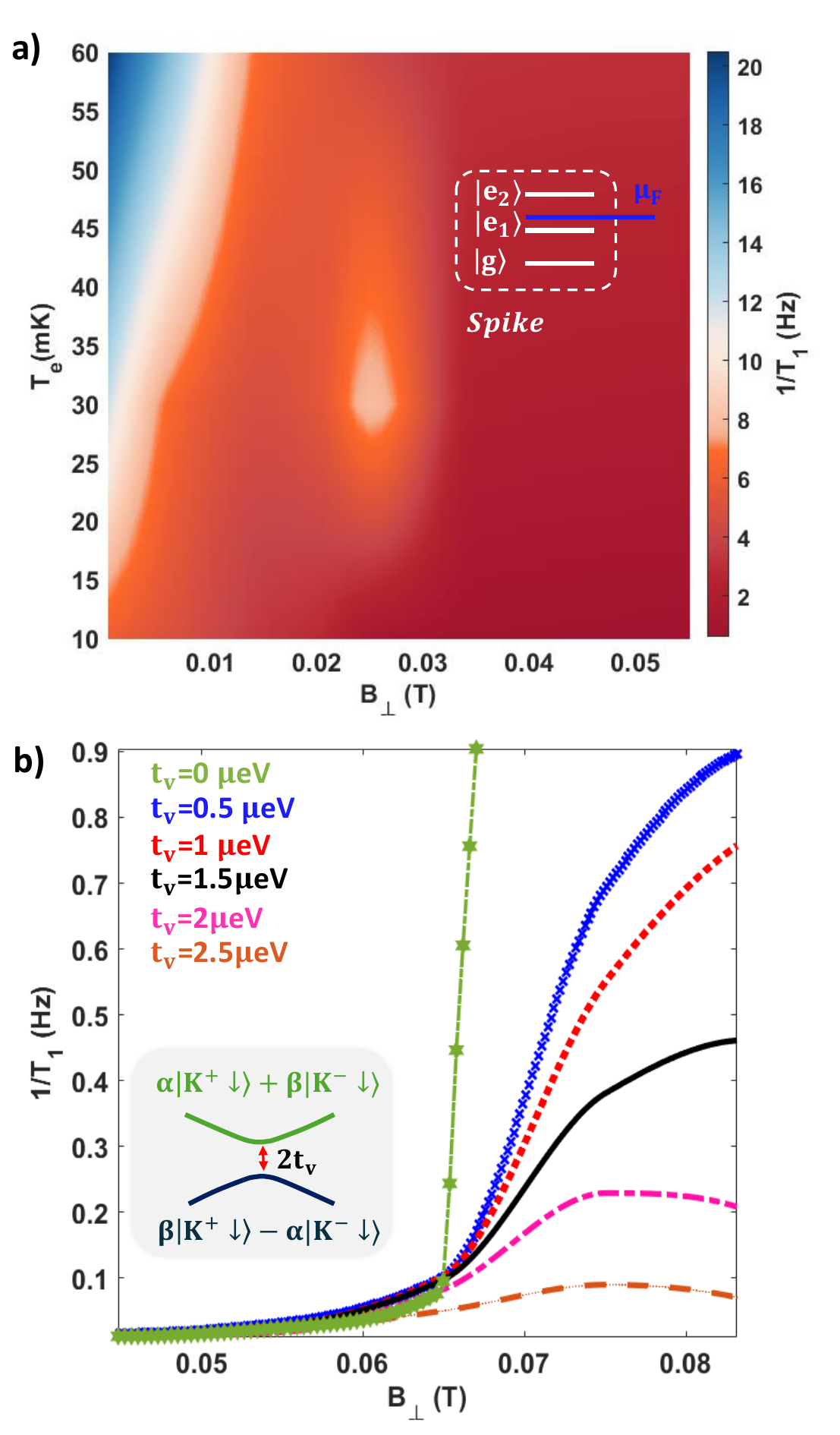}
\caption{Effect of $T_e$ and $t_v$ on the $1/T_1$ spikes. (a) Simulated $1/T_1$ versus $T_e$ at the spin/valley operating point where a spike is observed, which is not a typical signature. The $T_e$ sweeps confirm a smooth increase of $1/T_1$ without any spike if the potential fluctuation is not included. (b) The $1/T_1$ versus $B_{\perp}$ around the spin--valley anticrossing. In this region, the sharp rise in $1/T_1$ is governed by $t_v$. 
Spectral constraints result in $t_v \lesssim 2~\mu\mathrm{eV}$.}
\label{fig:Fig4}
\end{figure}
\indent This mechanism is incorporated in the simulations by initializing population in both $\bar{2}$ and $\bar{3}$, weighted by the probability of charge-offset events. The resulting fast initial decay ($\gamma_v$) and slow tail ($\gamma_k$) arise from the distinct decay pathways shown in the inset state-transition diagrams. Enforcing detailed balance, $\gamma_{ji}/\gamma_{ij}=e^{-(E_j-E_i)/k_{\mathrm{B}}T_e}$, the simulations reproduce both decay components, confirming that the observed scenario originates from the loading dynamics.\\ 
\indent {\textit{Consolidating the results.}} We observe that $\gamma_v$ ($\gamma_{\bar{3}\bar{2}}$ around the anticrossing) is strongly modulated by mixing and exhibits a pronounced $B_\perp$ dependence, as shown in Fig.~\ref{fig:Fig2}. Since $\gamma_s$ has a weak $B_{\perp}$ dependence, the variations in the relaxation trends observed for the spin/valley protocol in the field range $0\text{--}55~\mathrm{mT}$ are primarily due to $\gamma_v$.
\begin{figure}[h]
\centering
\includegraphics[width=0.85\columnwidth]{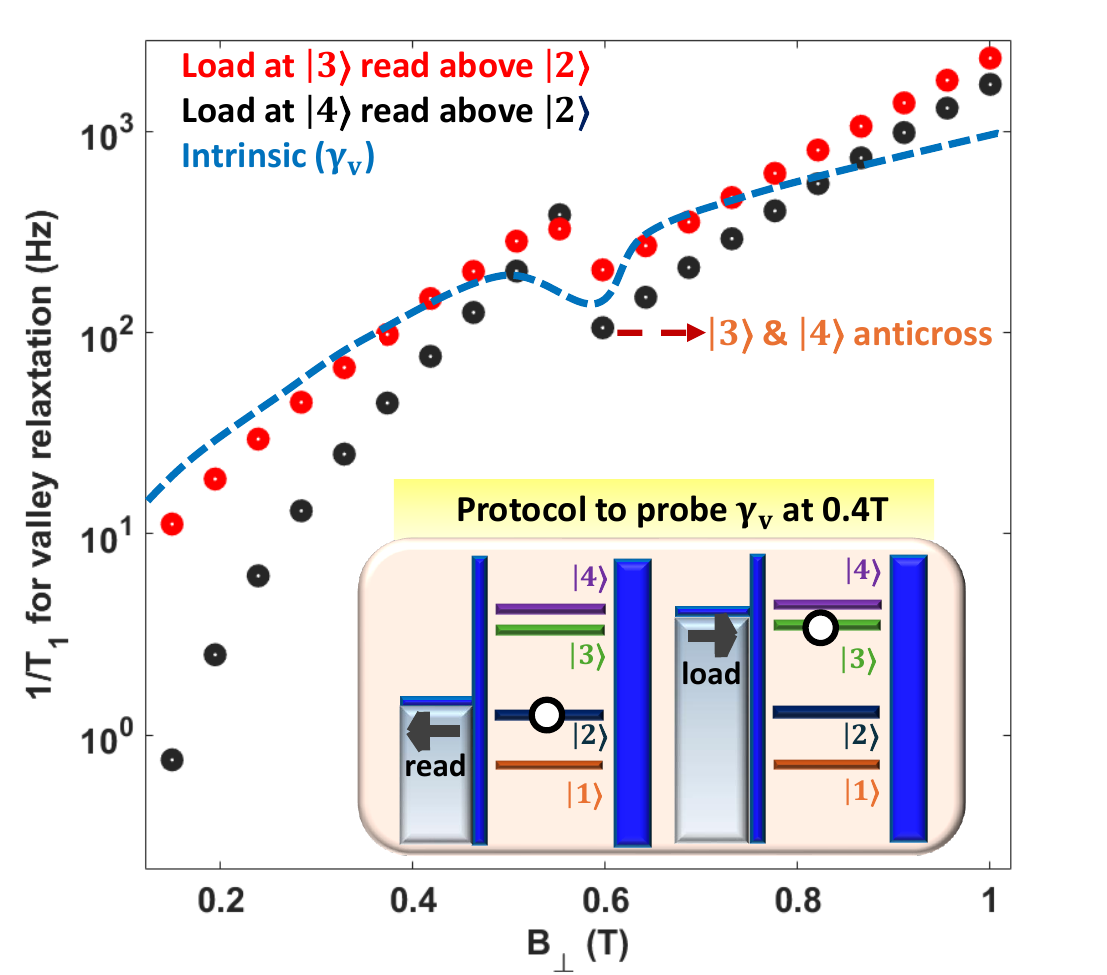}
\caption{Designing readouts. Predicted valley $1/T_{1}$ versus $B_{\perp}$ upto $1~\mathrm{T}$. The $B_{\perp}$ dependence follows the $\gamma_v$ with a weak non-monotonic shoulder near the second anticrossing due to $t_v$. The inset depicts the readout protocol.}
\label{fig:Fig5}
\end{figure}
Accounting for multilevel dynamics, the variation of the rates observed in the $80\text{--}100~\mathrm{mT}$ range (yellow patch in Fig.~\ref{fig:Fig2}) is also dominated by valley flips.\\
\indent The spike in the spin/valley data around $B_{\perp}\!\approx\!25~\mathrm{mT}$ is reproduced by incorporating the probability of sudden dot-potential fluctuations, which transiently shift $\mu_F$ relative to the loading window~\cite{keith2019benchmarking,garreis2023counting}. To substantiate this interpretation, $T_e$ is used as an independent control parameter. As shown in Fig.~\ref{fig:Fig4}(a), increasing $T_e$ uniformly degrades $T_1$ while preserving the $B_{\perp}$ dependence, thereby ruling out thermal activation as the origin of the spike.\\
\indent In the intermediate field range $55\text{--}78~\mathrm{mT}$ (red patch in Fig.~\ref{fig:Fig2}), variations in $\gamma_{\bar{2}1}$ become most pronounced, reflecting the modulation of $\gamma_k$. Despite its small magnitude, $t_v$ plays a decisive role in shaping the relaxation trend, as demonstrated in Fig.~\ref{fig:Fig4}(b). 
Importantly, the mismatch in the Kramers lifetime estimate highlighted in Fig.~\ref{fig:Fig2} (approximately $4~\mathrm{s}$) cannot be attributed solely to the variations in $t_v$ or to the linear combinations of $\gamma_{32}$, $\gamma_s$, and $\gamma_{21}$. The effective rate $\gamma_{\mathrm{effective}}$ is only reproduced via the integrated framework.\\
\indent Figure~\ref{fig:Fig5} outlines a protocol for directly probing $\gamma_v$ by suppressing spin-flip pathways. Away from the anticrossing (See Supplementary Note III), this protocol will recover the intrinsic valley $T_1$ time. Near the anticrossing however, mixing between the $\bar{\lvert 3\rangle}$ and $\bar{\lvert 4\rangle}$ states modifies the accessible decay pathways, such that the experimentally extracted lifetime can no longer be directly related to the intrinsic $T_1$ time.\\
\indent \textit{Conclusion.} We developed a generalized two-tier approach to provide a consistent interpretation of lifetime estimates from multilevel ERT \cite{denisov2025spin} experiments. By reproducing recent experimental trends \cite{denisov2025spin}, we made a clear distinction between the QTS $T_1$ time and the effective lifetime. Near degeneracies and anticrossings, one has to resort to the effective lifetime - which is a combination of both intrinsic and extrinsic contributions. We also pointed out that the generalized Matthiessen’s rule in such cases, can also be violated. Near zero fields, 
one may record hours-long $T_{1}$ time if the system is tuned to probe Kramers transitions, reflecting the strong symmetry protection of Kramers pairs. The integrated approach developed here can be employed to explain many existing discrepancies between intrinsic estimates and measured lifetimes \cite{banszerus2021spin,banszerus2025phonon,garreis2024long} as well as advancing further experiments in emerging qubit systems.\\
\indent \textit{Acknowledgments.} We acknowledge useful discussions with Lin Wang. BM acknowledges funding from the the Inani Chair Professorship fund through Grant No.~DO/2024-INAN/001-001, and the Department of Science and Technology, Government of India, under the National Quantum Mission through Grant no.~DST/QTC/NQM/QMD/2024/4. BM and SK acknowledge funding from the Dhananjay Joshi Endowment award through Grant No: DO/2023-DJEF002. SK acknowledges the Swasth Research Fellowship. AM acknowledges the Prime Minister Research Fellowship (PMRF). BW acknowledges support from the Singapore Ministry of Education (MOE) Academic Research Fund Tier 3 grant (MOE-MOET32023-0003) “Quantum Geometric Advantage” and the Air Force Office of Scientific Research under award number FA2386-24-1-4064.
GB acknowledges funding from the Deutsche Forschungsgemeinschaft (DFG, German
Research Foundation) – Project No. 425217212 – SFB 1432. 
\bibliography{references}
\onecolumngrid
\section*{Supplementary Material}
\section{Calculation of intrinsic rates}
\subsection*{Valley relaxation}
\emph{\textbf{Phonon noise}}: \indent Dot eigenstates in valley $\tau \in \{K^{+},K^{-}\}$ (with spin acting as a spectator) are described by two-component envelope functions~\cite{recher2009bound}. The eigenstates take the form \\
\begin{equation}
\Psi^{(\tau)}_{nj}(r,\theta)
=
\frac{1}{\sqrt{2\pi}}
\begin{pmatrix}
F_{nj}(r)e^{i(j-1)\theta} \\
G_{nj}(r)e^{ij\theta}
\end{pmatrix},
\end{equation}
with the normalization condition, \\
\begin{equation*}
\int_{0}^{\infty} 2\pi r\,dr
\left(
|F_{nj}(r)|^2 + |G_{nj}(r)|^2
\right)
=1.
\end{equation*}
\indent The in-plane acoustic electron–phonon coupling (EPC) Hamiltonian in the bilayer graphene (BLG) sublattice basis can be written as~\cite{ando2005theory,wang2024valley} \\
\begin{equation}
H^{\lambda q}_{\mathrm{EPC}}
=
\frac{q}{\sqrt{A\rho\Omega_{\lambda q}}}
\left[
g_1 a_1(\lambda)
+
g_2 a_2(\lambda,\phi)\sigma_x
\right]
\left(
e^{i\mathbf{q}\cdot\mathbf{r}} b_{\lambda q}
-
e^{-i\mathbf{q}\cdot\mathbf{r}} b^\dagger_{\lambda q}
\right),
\end{equation}
\indent where $\Omega_{\lambda q}=v_\lambda q$ is the phonon frequency for branch $\lambda$, $A$ is the sample area, and $\rho$ is the mass density. The coefficients $g_1$ and $g_2$ correspond to the deformation potential (DP) and bond-length change (BLC) coupling mechanisms, respectively. The polarization factors satisfy $a_1=i$ for longitudinal acoustic (LA) phonons and $a_1=0$ for transverse acoustic (TA) phonons, indicating that the DP couples only to LA phonons. The BLC coupling enters through $a_2(\lambda,\phi)$, with $a_2=ie^{2i\phi}$ for LA modes and $a_2=i$ for TA modes.\\
\indent Acoustic phonons in BLG carry small momenta ($q \ll K$), and therefore direct intervalley scattering between the two valleys $K^{+}$ and $K^{-}$ is strongly suppressed. As a result, the electron–phonon matrix elements are evaluated within a single valley basis, and the valley eigenstates are subsequently combined through the appropriate mixing coefficients. To evaluate the matrix elements, we expand the plane-wave factor as,

\begin{equation*}
e^{i\mathbf{q}\cdot\mathbf{r}}
=
\sum_{m=-\infty}^{\infty}
i^{m}
J_m(qr)
e^{im(\theta-\phi)} .
\end{equation*}
\indent From this expansion the angular momentum selection rule follows directly \\
\begin{equation*}
m=j'-j\equiv \Delta j .
\end{equation*}
\indent At small phonon momenta the leading nonvanishing contribution corresponds to the dipole channel with $|\Delta j|=1$. In this limit the Bessel function reduces to $J_{\pm1}(qr) \simeq \frac{qr}{2}.$\\
\indent The relevant radial overlap integrals then become
\begin{equation}
M_{n'n}
=
\int_{0}^{\infty} r^2 dr
\left[
F_{n'j\pm1}(r)F_{nj}(r)
+
G_{n'j\pm1}(r)G_{nj}(r)
\right],
\end{equation}
\begin{equation}
N^{AB}_{n'n}
=
\int_{0}^{\infty} r^2 dr
\left[
F_{n'j\pm1}(r)G_{nj}(r)
+
G_{n'j\pm1}(r)F_{nj}(r)
\right].
\end{equation}

\indent Here $M_{n'n}$ describes intra-sublattice processes associated with the DP coupling, while $N^{AB}_{n'n}$ corresponds to inter-sublattice transitions arising from BLC. \\
\indent The corresponding matrix elements of the plane-wave operator are therefore
\begin{equation*}
\langle n'j\pm1|e^{i\mathbf{q}\cdot\mathbf{r}}|nj\rangle_{\mathrm{DP}}
\simeq
\pi q e^{\mp i\phi} M_{n'n},
\end{equation*}
\begin{equation*}
\langle n'j\pm1|e^{i\mathbf{q}\cdot\mathbf{r}}|nj\rangle_{\mathrm{BLC}}
\simeq
\pi q e^{\mp i\phi} N^{AB}_{n'n}.
\end{equation*}
\indent Including the EPC prefactor 
$
\frac{q}{\sqrt{A\rho\Omega_{\lambda q}}}
=
\frac{\sqrt{q}}{\sqrt{A\rho v_\lambda}}$, the one-phonon transition amplitudes become \\
\begin{equation}
A^{(\mathrm{DP,LA})}_{n'j\pm1,nj}
=
\frac{\sqrt{q}}{\sqrt{A\rho v_{\mathrm{LA}}}}
g_1 a_1(\mathrm{LA})
(\pi q)
e^{\mp i\phi}
M_{n'n},
\end{equation}
\begin{equation}
A^{(\mathrm{BLC},\lambda)}_{n'j\pm1,nj}
=
\frac{\sqrt{q}}{\sqrt{A\rho v_\lambda}}
g_2 a_2(\lambda,\phi)
(\pi q)
e^{\mp i\phi}
N^{AB}_{n'n}.
\end{equation}
\indent The phonon-induced relaxation rate is obtained from Fermi’s golden rule (FGR) \\
\begin{equation}
\gamma_{n\rightarrow n'}
=
\frac{2\pi}{\hbar}
\sum_{\lambda,q}
\left|
A^{(\lambda)}_{n'n}
\right|^2
\delta(\Delta E-\hbar\omega_{\lambda q}).
\end{equation}
\indent For a two-dimensional (2D) phonon bath the momentum sum is replaced by
$\sum_q \rightarrow
A \int \frac{dq\,q}{2\pi}$.\\
Performing the density-of-states (DOS) integration yields a power-law dependence on the phonon momentum $\gamma \propto q^4$.\\
\indent The valley splitting is determined by the valley Zeeman energy 
$
\Delta E = g_v \mu_B B_\perp,
$ which fixes the phonon momentum through the energy conservation condition 
$
q = \frac{\Delta E}{\hbar v_\lambda}
\propto B_\perp.
$
\\ \indent The valley relaxation rate therefore becomes,
\begin{equation}
\gamma_v(B_\perp)
=
\frac{
g_1^2 |M_{n'n}|^2
+
g_2^2 |N^{AB}_{n'n}|^2
}{
\rho v_\lambda^6 \hbar^5
}
(g_v\mu_B B_\perp)^4 .
\end{equation}
\indent Thus the valley relaxation rate in BLG due to acoustic phonons follows the quartic magnetic-field scaling \\
\begin{equation}
\gamma_v(B_\perp)
\propto
\left(
g_1^2 |M_{n'n}|^2
+
g_2^2 |N^{AB}_{n'n}|^2
\right)
B_\perp^4,
\end{equation}
which originates from the dipole-like selection rule $|\Delta j|=1$ and the linear dispersion of acoustic phonons. Here the DP coupling $g_1$ contributes only through LA phonons, whereas the BLC coupling $g_2$ contributes to both LA and TA phonon branches.\\
\indent \emph{\textbf{Phonon noise}}: In addition to phonon-mediated processes, valley relaxation in BLG quantum dots (QD) can also be induced by fluctuating electric fields originating from Johnson noise in the surrounding circuit. These electric-field fluctuations couple to the electric dipole of the confined electron and provide an additional relaxation channel between valley states. The corresponding system–bath interaction Hamiltonian is,
\begin{equation}
H_{\mathrm{int}}^{J} = -e\,\mathbf{r}\cdot\mathbf{E}(t),
\end{equation}
where $\mathbf{E}(t)$ represents the electric field fluctuations arising from voltage noise in the surrounding circuit. \\
\indent Using FGR, the transition rate between valley states can be written as, \\
\begin{equation}
\gamma_v^{J}
=
\frac{1}{\hbar^2}
\int_{-\infty}^{\infty}
d\tau
\,e^{i\omega_v\tau}
\langle
E_\alpha(\tau)E_\alpha(0)
\rangle
|\langle n'|e r_\alpha|n\rangle|^2,
\end{equation}
where $\omega_v=\Delta E/\hbar$ denotes the valley transition frequency and $\alpha$ labels the in-plane Cartesian components of the electric field.\\
\indent The electric-field fluctuations are related to voltage fluctuations across the circuit impedance $R$. The spectral density of voltage noise is given by,
\begin{equation}
S_V(\omega)
=
2R\hbar\omega
\coth
\left(
\frac{\hbar\omega}{2k_B T}
\right).
\end{equation}
\indent In the low-temperature limit ($\hbar\omega \gg k_B T$), the thermal factor approaches unity and the spectrum reduces to the quantum limit, 
$
S_V(\omega) \simeq 2R\hbar\omega .
$

\indent The electric-field spectral density is related to the voltage noise through the characteristic dot–gate coupling length $l$, \\
\begin{equation}
S_E(\omega)
\simeq
\frac{S_V(\omega)}{l^2}
=
\frac{2R\hbar\omega}{l^2}.
\end{equation}
\indent The electric field couples to the dipole moment associated with the valley degree of freedom. The dipole matrix element between valley eigenstates can therefore be written as, $
|\langle n'|e\mathbf{r}|n\rangle|^2
=
e^2 d_v^2,
$ where $d_v$ denotes the effective valley dipole length arising from the spatial structure of the valley eigenstates. In electrostatically defined BLG QDs, the valley eigenstates are not perfectly orthogonal in real space due to the finite confinement potential and the presence of intervalley mixing. This generates a small but finite electric dipole moment $d_v$, which enables coupling between valley states and fluctuating electric fields. \\
\indent Substituting these expressions into FGR yields,
\begin{equation}
\gamma_v^{J}(\omega_v)
=
\frac{1}{\hbar^2}
S_E(\omega_v)
e^2 d_v^2 .
\end{equation}
\indent Using $\omega_v = g_v\mu_B B_\perp/\hbar$, the magnetic-field dependence of the relaxation rate becomes, \\
\begin{equation}
\gamma_v^{J}(B_\perp)
=
\frac{2R}{\hbar^2 l^2}
e^2 d_v^2
(g_v\mu_B B_\perp).
\end{equation}
\indent Thus the Johnson-noise-mediated valley relaxation rate scales linearly with $B_\perp$,
in contrast to the quartic magnetic-field dependence of the phonon-mediated valley relaxation derived above. The prefactor depends on the circuit resistance $R$, the electrostatic length scale $l$, and the effective valley dipole length $d_v$ characterizing the coupling between the valley states and the fluctuating electric field.
\subsection*{Spin relaxation}
\emph{\textbf{Phonon noise:}}
Spin relaxation in BLG QDs arises primarily through the admixture mechanism, in which spin--orbit coupling (SOC) mixes opposite-spin states within the orbital spectrum of the dot. A weak SOC Hamiltonian $H_{\mathrm{SO}}$, admixes the spin state $|0,\uparrow\rangle$ with excited states (ES) $|n,\downarrow\rangle$. The strength of this admixture is quantified by

\begin{equation*}
\eta_n =
\frac{\langle n,\downarrow | H_{\mathrm{SO}} | 0,\uparrow \rangle}
{E_0 - E_n}
\propto \Delta_R ,
\end{equation*}
where $\Delta_R$ denotes the Rashba spin splitting.

Because the electron–phonon interaction conserves spin, a spin flip cannot occur through EPC alone. The leading contribution therefore arises from a second-order process involving both the SO interaction and the electron–phonon interaction. The corresponding transition amplitude can be written as

\begin{equation}
M^{(\lambda q)}_{\downarrow\uparrow}
=
\sum_{n\neq0}
\left[
\frac{
\langle0,\downarrow|H^{\lambda q}_{\mathrm{EPC}}|n,\downarrow\rangle
\langle n,\downarrow|H_{\mathrm{SO}}|0,\uparrow\rangle
}{E_0-E_n}
+
\frac{
\langle0,\downarrow|H_{\mathrm{SO}}|n,\uparrow\rangle
\langle n,\uparrow|H^{\lambda q}_{\mathrm{EPC}}|0,\uparrow\rangle
}{E_0-E_n}
\right].
\end{equation}

The energy splitting between the two spin states is determined by the spin Zeeman energy, $\Delta E = g_s \mu_B B_\perp.$
The emitted acoustic phonon therefore satisfies $\hbar\omega=\Delta E$.

At low $B_\perp$ the transition amplitude reduces to an admixture prefactor proportional to $\Delta_R$ multiplied by an orbital EPC matrix element. The angular dependence of the resulting transition probability is captured by

\begin{equation}
f(\theta)
=
\cos^4\!\left(\frac{\theta}{2}\right)
+
\sin^4\!\left(\frac{\theta}{2}\right)
=
\frac{3+\cos(2\theta)}{4},
\end{equation}

which arises from the angular structure of the dipole matrix elements associated with the phonon propagation direction.

To evaluate the orbital matrix elements we expand the plane-wave factor for small phonon momenta,
\begin{equation*}
e^{i\mathbf q\cdot\mathbf r}
\simeq
1 + i\mathbf q\cdot\mathbf r .
\end{equation*}

Defining the dipole matrix elements by $\mathbf r_{n0} = \langle n | \mathbf r | 0 \rangle$, and using the angular momentum selection rule $\Delta j=\pm1$, the EPC matrix elements take the form
\begin{equation*}
\langle0|H^{\mathrm{LA,DP}}_{\mathrm{EPC}}|n\rangle
\propto
g_1
\frac{q}{\sqrt{\Omega_{\mathrm{LA}q}}}
(i\,\mathbf q\cdot\mathbf r_{n0}),
\end{equation*}
which leads to
\begin{equation*}
|\langle0|H|n\rangle|^2
\propto
g_1^2
\frac{q^4}{\Omega_{\mathrm{LA}q}}
=
g_1^2\frac{q^3}{v_{\mathrm{LA}}}.
\end{equation*}

Similarly, for BLC coupling mechanisms, we obtain,
\begin{equation*}
\langle0|H^{\mathrm{BLC}}_{\mathrm{EPC}}|n\rangle
\propto
g_2
\frac{1}{\sqrt{\Omega_{\lambda q}}}
(i\,\mathbf q\cdot\mathbf r_{n0}),
\end{equation*}
which yields
\begin{equation*}
|\langle0|H|n\rangle|^2
\propto
g_2^2
\frac{q^2}{\Omega_{\lambda q}}
=
g_2^2\frac{q}{v_\lambda}.
\end{equation*}

Applying FGR, the deformation-DP contribution to the spin relaxation rate becomes

\begin{equation}
\gamma^{\lambda,\mathrm{DP}}_{s}(B_\perp)
=
\frac{C_{\mathrm{DP}} g_1^2}{\rho}
\frac{(\Delta E)^4}{v_{\mathrm{LA}}^6}
f(\theta)
\left|
\sum_{n'\neq n}
M_{n'n}R_{n'n}
\left(
\delta_{j,j+1}N^{AB}_{n'n}
+
\delta_{j,j-1}N^{AB}_{n'n}
\right)
\right|^2 ,
\end{equation}
where $C_{\mathrm{DP}}$ is a numerical prefactor and $M_{n'n}$ and $R_{n'n}$ denote orbital overlap integrals.

The BLC mechanism yields

\begin{equation}
\gamma^{\lambda,\mathrm{BLC}}_{s}(B_\perp)
=
\frac{C^{\lambda}_{\mathrm{BLC}} g_2^2}{\rho}
\frac{(\Delta E)^2}{v_\lambda^4}
f(\theta)
\left|
\sum_{n'\neq n}
R_{n'n}
\left[
\delta_{j,j+1}(N^{AB}_{n'n})^2
+
\delta_{j,j-1}(N^{AB}_{n'n})^2
\right]
\right|^2 .
\end{equation}

These power-law scalings arise from the combined effect of the SOC-induced admixture amplitude, the dipole-like orbital matrix element, and the linear dispersion of in-plane acoustic phonons.

In contrast, direct spin--phonon coupling to out-of-plane flexural (ZA) phonons, which can arise in graphene through curvature-induced spin--orbit interactions, is not operative in the hBN-encapsulated device geometry considered here. In electrostatically defined BLG QDs within such gate stacks, flexural distortions are strongly suppressed, and the dominant spin relaxation channels therefore originate from the in-plane acoustic phonon mechanisms derived above.

\emph{\textbf{Johnson noise:}}
In addition to phonon-mediated processes, fluctuating electric fields generated by Johnson noise in the surrounding circuit provide an additional relaxation channel. These electric-field fluctuations couple to the dipole moment induced by SOC admixture and can drive spin transitions. The corresponding relaxation rate can be written as
\begin{equation}
\gamma_s^{J}(B_\perp)
=
\frac{2R}{\hbar^2 l^2}
e^2 d_s^2 \Delta E,
\end{equation}
where $R$ is the effective circuit resistance, $l$ denotes the dot--gate distance, and $d_s$ is the effective dipole length associated with the SOC-induced spin admixture. Using $\Delta E=g_s\mu_B B_\perp$, this becomes
\begin{equation}
\gamma_s^{J}(B_\perp)
=
\frac{2R}{\hbar^2 l^2}
e^2 d_s^2
(g_s\mu_B B_\perp).
\end{equation}

Thus the Johnson-noise-mediated spin relaxation rate scales linearly with the $B_\perp$.
\\
\emph{\textbf{Absence of Van Vleck cancellation for spin and valley qubits in BLG:}}
The leading-order spin--phonon matrix elements do not cancel at zero magnetic field because the two spin states involved in the transition are not related by time reversal. Consequently, the Van Vleck cancellation that suppresses relaxation between Kramers partners is absent for both valley and spin mediated transitions and the relaxation rates retain the lowest-order $B_\perp$ scaling derived above.
\subsection*{Kramers relaxation}
\indent \emph{\textbf{Phonon noise}}: Kramers partners within a single dot orbital $n$ are related by time reversal $\mathcal{T}$, \\
\[
|i\rangle = |n, K^{+},\uparrow_B\rangle, 
\qquad 
|f\rangle = \mathcal{T}|i\rangle = |n, K^{-},\downarrow_B\rangle .
\]
\indent The $H_{\mathrm{EPC}}$ is time-reversal even, while the spin--orbit interaction $H_{\mathrm{SO}}$ admixes opposite-spin virtual states. The lowest nonzero one-phonon process is therefore second order, involving one EPC vertex and one SOC vertex. The corresponding effective transition amplitude is \\
\begin{equation*}
M_{\mathrm{eff}}(\mathbf{q})
=
\sum_{m}
\frac{\langle f|H_{\mathrm{EPC}}|m\rangle \langle m|H_{\mathrm{SO}}|i\rangle}
{E_i-E_m-\tfrac{1}{2}E_Z}
+
\sum_{m}
\frac{\langle f|H_{\mathrm{SO}}|m\rangle \langle m|H_{\mathrm{EPC}}|i\rangle}
{E_i-E_m+\tfrac{1}{2}E_Z},
\end{equation*}
\indent where $E_Z=g_s\mu_B B_\perp$ and the sum runs over orbital excitations $|m\rangle$ with the appropriate valley structure. \\
\indent Let $\Delta_m=(E_n-E_m)_{B_\perp=0}$. Expanding the denominators for small $E_Z=\hbar\omega$ gives, \\
\begin{equation*}
\frac{1}{\Delta_m\mp E_Z/2}
=
\frac{1}{\Delta_m}
\pm
\frac{E_Z}{2}\frac{1}{\Delta_m^2}
+
O(E_Z^2).
\end{equation*}
\indent Define the time-reversal-related matrix elements
\begin{equation*}
A_m
=
\langle f|H_{\mathrm{EPC}}|m\rangle
\langle m|H_{\mathrm{SO}}|i\rangle,
\qquad
B_m
=
\langle f|H_{\mathrm{SO}}|m\rangle
\langle m|H_{\mathrm{EPC}}|i\rangle .
\end{equation*}
\indent For a Kramers pair and time-reversal-even $H_{\mathrm{EPC}}$, one has $B_m=-A_m$ at $B_\perp=0$. Consequently, the $E_Z^0$ term cancels exactly (Van Vleck cancellation), and the leading contribution is linear in $E_Z$, \\
\begin{equation*}
M_{\mathrm{eff}}(\mathbf{q})
=
\frac{E_Z}{2}\sum_m\frac{A_m-B_m}{\Delta_m^2}
+
O(E_Z^2)
=
E_Z\sum_m\frac{A_m}{\Delta_m^2}
+
O(E_Z^2)
\propto \omega .
\end{equation*}
\indent Thus, unlike the spin and valley channels, the effective Kramers transition amplitude acquires one explicit power of $\omega$ purely from Van Vleck cancellation. \\
\indent For long-wavelength acoustic phonons, we use the dipole limit
$e^{i\mathbf{q}\cdot\mathbf{r}}\simeq 1+i\,\mathbf{q}\cdot\mathbf{r}$ between different orbitals, with the selection rule $\Delta j=\pm1$. The EPC vertices may then be written as \\
\begin{equation*}
V_{fm}^{(\lambda)}
=
\frac{q}{\sqrt{A\rho \Omega_{\lambda q}}}\,
(\mathbf{q}\cdot \mathbf{d}_{0m}^{(\lambda)}),
\qquad
\Omega_{\lambda q}=v_\lambda q ,
\end{equation*}
where $\mathbf{d}_{0m}^{(\lambda)}$ denotes the dipole overlap for phonon branch $\lambda$. After angular averaging in two dimensions,
\begin{equation*}
\overline{|\hat{\mathbf q}\cdot \mathbf d|^2}_{\phi}
=
\frac12 |\mathbf d|^2 .
\end{equation*}
\indent We collect the phonon channels through the effective coupling constants
\[
|g_\lambda|^2=
\begin{cases}
g_1^2, & \text{(LA, DP)},\\
g_2^2, & \text{(LA, BLC) or (TA, BLC)},\\
0, & \text{(TA, DP)},
\end{cases}
\]
and define the radial dipole overlaps, consistent with $\Delta j=\pm1$, \\
\begin{equation*}
|d^{(\mathrm{DP})}_{0m}|^2 \to |M_{m0}|^2,
\qquad
M_{m0}
=
\int_{0}^{\infty}dr\, r^2\,[F_mF_0+G_mG_0],
\end{equation*}
\begin{equation*}
|d^{(\mathrm{BLC})}_{0m}|^2 \to |N^{AB}_{m0}|^2,
\qquad
N^{AB}_{m0}
=
\int_{0}^{\infty}dr\, r^2\,[F_mG_0+G_mF_0].
\end{equation*}
\indent Using Fermi’s golden rule, the phonon-induced Kramers relaxation rate is \\
\begin{equation*}
\gamma^{(\lambda)}
=
\frac{2\pi}{\hbar}
\frac{A}{(2\pi)^2}
\int d^2q\,
\delta(\hbar\Omega_{\lambda q}-\hbar\omega)\,
\overline{|M_{\mathrm{eff}}(\mathbf{q})|^2}_{\phi}.
\end{equation*}
\indent Together with
\[
\int q\,dq\, q^3\,\delta(\hbar v_\lambda q-\hbar\omega)
=
\frac{q_0^4}{\hbar v_\lambda},
\qquad
q_0=\frac{\omega}{v_\lambda},
\]
and the linear-in-$E_Z$ effective amplitude above, one obtains the characteristic $B_\perp^6$ scaling for Kramers relaxation, \\
\begin{equation*}
\gamma_{\mathrm{k}}^{(\lambda)}
\propto
\frac{\omega^6}{v_\lambda^6}
\sum_m
\frac{|\langle m|H_{\mathrm{SO}}|i\rangle|^2}{\Delta_m^4}
\times
\begin{cases}
g_1^2 |M_{m0}|^2, & \text{(LA, DP)},\\[2pt]
g_2^2 |N^{AB}_{m0}|^2, & \text{(LA/TA, BLC)}.
\end{cases}
\end{equation*}
\indent Restoring prefactors, the explicit rates read \\
\begin{equation}
\gamma_{\mathrm{k}}^{(\mathrm{LA,DP})}
=
\frac{g_1^{2}}{4\pi\rho}\,
\frac{\omega^{6}}{v_{\mathrm{LA}}^{6}}\;
\sum_{m}
\frac{|\langle m|H_{\mathrm{SO}}|i\rangle|^{2}}{\Delta_{m}^{4}}\;
|M_{m0}|^{2},
\end{equation}
\begin{equation}
\gamma_{\mathrm{k}}^{(\mathrm{LA,BLC})}
=
\frac{g_2^{2}}{4\pi\rho}\,
\frac{\omega^{6}}{v_{\mathrm{LA}}^{6}}\;
\sum_{m}
\frac{|\langle m|H_{\mathrm{SO}}|i\rangle|^{2}}{\Delta_{m}^{4}}\;
|N^{AB}_{m0}|^{2},
\end{equation}
\begin{equation}
\gamma_{\mathrm{k}}^{(\mathrm{TA,BLC})}
=
\frac{g_2^{2}}{4\pi\rho}\,
\frac{\omega^{6}}{v_{\mathrm{TA}}^{6}}\;
\sum_{m}
\frac{|\langle m|H_{\mathrm{SO}}|i\rangle|^{2}}{\Delta_{m}^{4}}\;
|N^{AB}_{m0}|^{2}.
\end{equation}
\indent The TA mode has no contribution from the DP channel to linear order. Here, $\omega=\frac{E_Z}{\hbar}=\frac{g_s\mu_B B_\perp}{\hbar}.$
The Van Vleck cancellation at $B_{\perp}=0$ eliminates the field-independent contribution to the transition amplitude. As a result, the leading term of the effective matrix element becomes proportional to the transition frequency $\omega$. When this frequency dependence is combined with the dipole coupling to phonons and the phase space available to $2$D acoustic phonons, the relaxation rate acquires a characteristic scaling proportional to $B_{\perp}^{6}$ for acoustic-phonon–induced relaxation between Kramers partners in BLG. \\
\indent \emph{\textbf{Johnson noise}}: In addition to phonon-mediated processes, fluctuating electric fields generated by Johnson noise in the surrounding circuit provide an additional relaxation channel for the Kramers qubit. The corresponding interaction Hamiltonian is \\
\begin{equation}
H_{\mathrm{int}}^{J}=-e\,\mathbf r\cdot \mathbf E(t).
\end{equation}
\indent For the Kramers pair
\[
|i\rangle = |n,K^{+},\uparrow_B\rangle,
\qquad
|f\rangle = \mathcal T|i\rangle = |n,K^{-},\downarrow_B\rangle,
\]
the direct dipole matrix element vanishes at $B_\perp=0$ because $H_{\mathrm{int}}^{J}$ is time-reversal even. The leading nonzero contribution therefore arises through second-order admixture with virtual ES, \\
\begin{equation}
D_{\mathrm{eff}}^{(\alpha)}
=
\sum_m
\frac{\langle f|e r_\alpha|m\rangle \langle m|H_{\mathrm{SO}}|i\rangle}
{E_i-E_m-\tfrac12 E_Z}
+
\sum_m
\frac{\langle f|H_{\mathrm{SO}}|m\rangle \langle m|e r_\alpha|i\rangle}
{E_i-E_m+\tfrac12 E_Z},
\end{equation}
where $E_Z=g_s\mu_B B_\perp$. Defining,
\[
A_m^{(\alpha)}=\langle f|e r_\alpha|m\rangle \langle m|H_{\mathrm{SO}}|i\rangle,
\qquad
B_m^{(\alpha)}=\langle f|H_{\mathrm{SO}}|m\rangle \langle m|e r_\alpha|i\rangle,
\]
and $\Delta_m=(E_n-E_m)_{B_\perp=0}$, one has $B_m^{(\alpha)}=-A_m^{(\alpha)}$ at $B_\perp=0$. Expanding for small $E_Z$ then gives \\
\begin{equation*}
D_{\mathrm{eff}}^{(\alpha)}
=
E_Z
\sum_m
\frac{A_m^{(\alpha)}}{\Delta_m^2}
+O(E_Z^2)
\propto B_\perp .
\end{equation*}
\indent Using Fermi’s golden rule, the Johnson-noise-induced Kramers relaxation rate is \\
\begin{equation}
\gamma_{\mathrm{k}}^{J}
=
\frac{1}{\hbar^2}
\sum_\alpha
S_{E_\alpha}(\omega)
\left|D_{\mathrm{eff}}^{(\alpha)}\right|^2,
\qquad
\omega=\frac{E_Z}{\hbar}.
\end{equation}
\indent In the low-temperature quantum Johnson-noise regime,
\[
S_E(\omega)=\frac{2R\hbar\omega}{l^2},
\]
so that,
\begin{equation}
\gamma_{\mathrm{k}}^{J}(B_\perp)
=
\frac{2R}{\hbar^3 l^2}
\left(g_s\mu_B B_\perp\right)^3
\sum_\alpha
\left|
\sum_m
\frac{\langle f|e r_\alpha|m\rangle \langle m|H_{\mathrm{SO}}|i\rangle}
{\Delta_m^2}
\right|^2.
\end{equation}
\indent Thus the Johnson-noise-mediated Kramers relaxation rate follows the scaling, $\gamma_{\mathrm{k}}^{J}(B_\perp)\propto B_\perp^3.
$
\section{Spin--valley mixing induced by intervalley coupling}
In Sec.~I we derived the intrinsic relaxation rates associated with transitions between the pure eigenstates of the BLG QD arising from electron–phonon and Johnson-noise coupling. Near spectral anticrossings, hybridization modifies the eigenbasis, and the observable transition rates between the hybridized states become weighted combinations of these intrinsic rates.

We first consider the anticrossing arising from intervalley coupling, parametrized by the coupling strength $t_v$.  In this regime the intervalley interaction mixes states belonging to the two valleys. The resulting eigenstates become superpositions of $K^+$ and $K^-$ valley components. The relevant eigenstates close to the anticrossing can therefore be written as
\begin{equation}
\begin{aligned}
|1\rangle &= |K^{-},\uparrow\rangle, \\
|\bar{2}\rangle &= \sin\!\left(\frac{\theta}{2}\right)|K^{+},\downarrow\rangle 
-\cos\!\left(\frac{\theta}{2}\right)|K^{-},\downarrow\rangle, \\
|\bar{3}\rangle &= \cos\!\left(\frac{\theta}{2}\right)|K^{+},\downarrow\rangle 
+\sin\!\left(\frac{\theta}{2}\right)|K^{-},\downarrow\rangle, \\
|4\rangle &= |K^{+},\uparrow\rangle .
\end{aligned}
\end{equation}

Here $\theta$ denotes the intervalley mixing angle determined by the competition between $t_v$ and the detuning between the uncoupled spin--valley states. Far from the anticrossing the eigenstates approach pure valley states, while near the anticrossing they become strongly hybridized.

Because the hybridized states contain contributions from multiple spin--valley configurations, transitions between them receive contributions from several intrinsic relaxation channels simultaneously. The intrinsic rates derived in Sec.~I correspond to
\begin{align*}
\gamma_s &: \text{spin flip within the same valley}, \\
\gamma_v &: \text{valley flip without spin flip}, \\
\gamma_k &: \text{combined spin--valley (Kramers) transition}.
\end{align*}
These intrinsic processes combine to produce the effective relaxation rates between the hybridized eigenstates.

\paragraph*{(i) Transition $\bar{2}\rightarrow1$.}

We evaluate
\begin{equation*}
\gamma_{\bar{2}1} = \left| \langle 1 | H_{\mathrm{e\mbox{-}ph}} | \bar{2} \rangle \right|^2 .
\end{equation*}

Using
\begin{equation*}
|\bar{2}\rangle = \sin\!\left(\tfrac{\theta}{2}\right)|K^+,\downarrow\rangle 
          - \cos\!\left(\tfrac{\theta}{2}\right)|K^-,\downarrow\rangle ,
\end{equation*}
the matrix element becomes
\begin{align*}
\langle 1 | H_{\mathrm{e\mbox{-}ph}} | \bar{2} \rangle
&= \sin\!\left(\tfrac{\theta}{2}\right) 
   \langle K^-,\uparrow | H_{\mathrm{e\mbox{-}ph}} | K^+,\downarrow \rangle \\
&\quad - \cos\!\left(\tfrac{\theta}{2}\right) 
   \langle K^-,\uparrow | H_{\mathrm{e\mbox{-}ph}} | K^-,\downarrow \rangle \\
&= \sin\!\left(\tfrac{\theta}{2}\right) M_k 
   - \cos\!\left(\tfrac{\theta}{2}\right) M_s .
\end{align*}
\begin{figure}[t]
\centering
\includegraphics[width=0.95\linewidth]{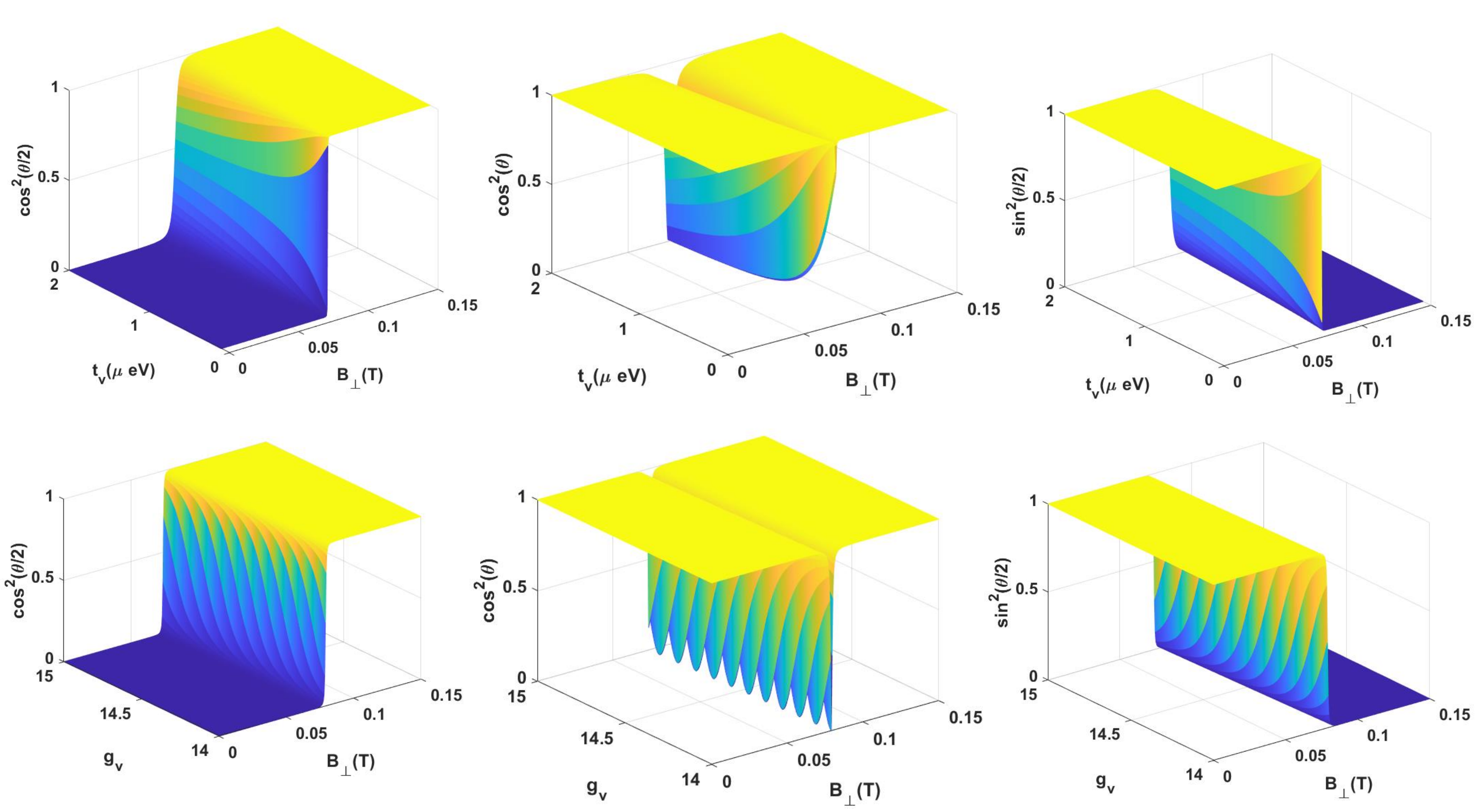}
\caption{Magnetic-field dependence of the state-mixing coefficients near the spin--valley anticrossing. 
At low and high $B_\perp$, the eigenstates remain nearly pure spin or valley, while near the anticrossing they strongly hybridize. 
The upper panels show the evolution of $\cos^2(\theta/2)$ and $\sin^2(\theta/2)$ as functions of intervalley coupling $t_v$ and perpendicular field $B_\perp$, 
while the lower panels illustrate the corresponding dependence on the valley $g$-factor $g_v$. 
The results highlight that both $t_v$ and $g_v$ critically tune the degree of hybridization at the anticrossing.}
\label{fig:Fig1_sup}
\end{figure}
The corresponding relaxation rates follow from FRG, 
$
\gamma = \frac{2\pi}{\hbar}\sum_{\mathbf{q},\lambda}
|M|^2
\delta(E_i-E_f-\hbar\omega_{\mathbf{q}\lambda}),
$
where $M=\langle f|H_{\mathrm{e-ph}}|i\rangle$ denotes the relevant electron--phonon matrix element. Accordingly, the intrinsic relaxation channels satisfy
\begin{align*}
\gamma_s &\propto |M_s|^2, \\
\gamma_k &\propto |M_k|^2, \\
\gamma_v &\propto |M_v|^2 .
\end{align*}

Squaring the matrix element gives
\begin{align*}
\left|\langle 1 | H_{\mathrm{e\mbox{-}ph}} | \bar{2} \rangle\right|^2
&= \left| \sin\!\left(\tfrac{\theta}{2}\right) M_k 
      - \cos\!\left(\tfrac{\theta}{2}\right) M_s \right|^2 \\
&= \sin^2\!\left(\tfrac{\theta}{2}\right)|M_k|^2 
 + \cos^2\!\left(\tfrac{\theta}{2}\right)|M_s|^2 \\
&\quad - 2\sin\!\left(\tfrac{\theta}{2}\right)\cos\!\left(\tfrac{\theta}{2}\right)
   \mathrm{Re}(M_k M_s^*).
\end{align*}

Assuming orthogonality of the phonon-induced matrix elements (no interference),
\begin{equation}
\gamma_{\bar{2}1} = \gamma_s \cos^2\!\left(\tfrac{\theta}{2}\right)
            + \gamma_k \sin^2\!\left(\tfrac{\theta}{2}\right).
\end{equation}

\paragraph*{(ii) Transition $\bar{3}\rightarrow1$.}

We evaluate
\begin{equation*}
\gamma_{\bar{3}1} = \left| \langle 1 | H_{\mathrm{e\mbox{-}ph}} | \bar{3} \rangle \right|^2 .
\end{equation*}

Using
\begin{equation*}
|\bar{3}\rangle = \cos\!\left(\tfrac{\theta}{2}\right)|K^+,\downarrow\rangle
          + \sin\!\left(\tfrac{\theta}{2}\right)|K^-,\downarrow\rangle ,
\end{equation*}
we obtain
\begin{align*}
\langle 1 | H_{\mathrm{e\mbox{-}ph}} | \bar{3} \rangle
&= \cos\!\left(\tfrac{\theta}{2}\right) M_k 
   + \sin\!\left(\tfrac{\theta}{2}\right) M_s .
\end{align*}

Squaring and neglecting interference terms gives
\begin{equation}
\gamma_{\bar{3}1} = \gamma_s \sin^2\!\left(\tfrac{\theta}{2}\right)
            + \gamma_k \cos^2\!\left(\tfrac{\theta}{2}\right).
\end{equation}

\paragraph*{(iii) Transition $\bar{3}\rightarrow\bar{2}$.}

This corresponds to a valley-flip transition between mixed down-spin states. Only matrix elements of the form
\begin{equation*}
\langle K^+,\downarrow | H_{\mathrm{e\mbox{-}ph}} | K^-,\downarrow \rangle = M_v
\end{equation*}
contribute. Expanding the hybridized states gives
\begin{align*}
\langle \bar{2} | H_{\mathrm{e\mbox{-}ph}} | \bar{3} \rangle
&= \sin\!\left(\tfrac{\theta}{2}\right)\cos\!\left(\tfrac{\theta}{2}\right)
   \Big(
   \langle K^+,\downarrow | H_{\mathrm{e\mbox{-}ph}} | K^-,\downarrow \rangle \\
&\qquad -
   \langle K^-,\downarrow | H_{\mathrm{e\mbox{-}ph}} | K^+,\downarrow \rangle
   \Big).
\end{align*}

By symmetry this reduces to the valley matrix element $M_v$, giving
\begin{equation}
\gamma_{\bar{3}\bar{2}} = \gamma_v \cos^2(\theta).
\end{equation}

\paragraph*{(iv) Transitions $4\rightarrow\bar{3}$ and $4\rightarrow\bar{2}$.}

These correspond to spin-flip processes from 
$|4\rangle = |K^+,\uparrow\rangle$ to the mixed down-spin states. The resulting rates are
\begin{align}
\gamma_{4\bar{3}} &= \gamma_s \cos^2\!\left(\tfrac{\theta}{2}\right) 
            + \gamma_k \sin^2\!\left(\tfrac{\theta}{2}\right), \\
\gamma_{4\bar{2}} &= \gamma_s \sin^2\!\left(\tfrac{\theta}{2}\right) 
            + \gamma_k \cos^2\!\left(\tfrac{\theta}{2}\right).
\end{align}

\paragraph*{(v) Transition $4\rightarrow1$.}
Finally, the transition between spin-up states in opposite valleys corresponds to a pure valley flip,
\begin{equation}
\gamma_{41} = \gamma_v .
\end{equation}
Thus, intervalley hybridization redistributes the intrinsic relaxation channels derived in Sec.~I, producing transition rates between hybridized states that depend explicitly on the mixing angle $\theta$, defined through $\cos^2\!\left(\frac{\theta}{2}\right)=\frac{\sqrt{(\Delta_{\mathrm{SO}}-E_Z)^2+4t_v^2}-(\Delta_{\mathrm{SO}}-E_Z)}{2\sqrt{(\Delta_{\mathrm{SO}}-E_Z)^2+4t_v^2}}$.
\section{Intra-valley spin mixing}
We now consider the anticrossing that appears at higher magnetic fields in the BLG spectrum, arising from spin mixing within a single valley. In contrast to the previous case, this anticrossing does not originate from 
intervalley coupling but instead arises from spin mixing within a single valley. 
Spin--orbit coupling generates an effective intra-valley spin-flip matrix element 
$t_s$ that hybridizes the states $|K^+,\uparrow\rangle$ and $|K^+,\downarrow\rangle$. 
Restricting to the $K^+$ subspace, the effective Hamiltonian can be written as
\begin{equation}
H_{K^+} =
\frac{1}{2}
\begin{pmatrix}
\Delta_{\mathrm{SO}} - E_Z & t_s \\
t_s & -(\Delta_{\mathrm{SO}} - E_Z)
\end{pmatrix},
\end{equation}
where $E_Z = g_s \mu_B B_\perp$ is the Zeeman energy. Diagonalizing this 
Hamiltonian yields the hybridized eigenstates given below,
\begin{align}
|1\rangle &= |K^{-},\uparrow\rangle,\\
|2\rangle &= |K^{-},\downarrow\rangle,\\
|\bar3\rangle &=
\sin\!\left(\frac{\phi}{2}\right)|K^{+},\uparrow\rangle
-
\cos\!\left(\frac{\phi}{2}\right)|K^{+},\downarrow\rangle,\\
|\bar4\rangle &=
\cos\!\left(\frac{\phi}{2}\right)|K^{+},\uparrow\rangle
+
\sin\!\left(\frac{\phi}{2}\right)|K^{+},\downarrow\rangle .
\end{align}

The mixing angle $\phi$ is determined by the detuning between the spin--orbit splitting $\Delta_{SO}$ and the Zeeman energy $E_Z$, together with the intra-valley spin-flip matrix element $t_s$. Explicitly,

\begin{align}
\cos^2\!\left(\frac{\phi}{2}\right)
&=
\frac{\sqrt{(\Delta_{SO}-E_Z)^2+4t_s^2}+(\Delta_{SO}-E_Z)}
{2\sqrt{(\Delta_{SO}-E_Z)^2+4t_s^2}},\\
\sin^2\!\left(\frac{\phi}{2}\right)
&=
\frac{\sqrt{(\Delta_{SO}-E_Z)^2+4t_s^2}-(\Delta_{SO}-E_Z)}
{2\sqrt{(\Delta_{SO}-E_Z)^2+4t_s^2}} .
\end{align}

Because the eigenstates are now spin-mixed within a single valley, the intrinsic relaxation channels derived in Sec.~I again combine to produce modified decay rates between the hybridized states.

Applying FGR yields
\begin{equation}
\begin{aligned}
\gamma_{21} &= \gamma_s,\\
\gamma_{\bar{3}1} &= \sin^2\!\left(\frac{\phi}{2}\right)\gamma_v
+
\cos^2\!\left(\frac{\phi}{2}\right)\gamma_k,\\
\gamma_{\bar{4}1} &= \cos^2\!\left(\frac{\phi}{2}\right)\gamma_v
+
\sin^2\!\left(\frac{\phi}{2}\right)\gamma_k,\\
\gamma_{\bar{3}2} &= \sin^2\!\left(\frac{\phi}{2}\right)\gamma_k
+
\cos^2\!\left(\frac{\phi}{2}\right)\gamma_v,\\
\gamma_{\bar{4}2} &= \cos^2\!\left(\frac{\phi}{2}\right)\gamma_k
+
\sin^2\!\left(\frac{\phi}{2}\right)\gamma_v,\\
\gamma_{\bar{4}\bar{3}} &= \sin^2\phi\,\gamma_s .
\end{aligned}
\end{equation}

The transition $\gamma_{21}$ corresponds to a pure spin flip in the $K^{-}$ valley and is therefore insensitive to the hybridization. The transitions $\gamma_{31}$ and $\gamma_{41}$ describe relaxation from the $K^{-}$ states to the hybridized $K^{+}$ states, with the relative contributions of valley-conserving and spin-assisted processes controlled by the mixing angle $\phi$.

Finally, the transition between the two hybridized $K^{+}$ states,
\[
\gamma_{\bar{4}\bar{3}}=\sin^2\phi\,\gamma_s ,
\]
represents intra-valley relaxation driven by spin mixing within the same valley. This process becomes significant only near the anticrossing where the hybridization is strongest, while far from the anticrossing the states revert to nearly pure spin eigenstates and the transition becomes strongly suppressed.
\section*{IV. Master-equation (ME) description of the load–wait dynamics}
\indent
We consider a BLG QD operating in the Coulomb-blockade (CB) regime and coupled to a single electronic reservoir. In the load/wait stage of the experimental cycle the tunnel barrier to the reservoir is closed, suppressing charge exchange with the lead. The remaining dynamics are governed by intrinsic noise-induced relaxation processes between well-separated energy eigenstates of the dot. Because these processes occur incoherently and rapidly destroy phase coherence between eigenstates, off-diagonal density-matrix elements can be neglected. The dynamics therefore reduce to classical population evolution, allowing the system to be described by a Pauli master equation for the state probabilities rather than a full quantum master equation. 

The relevant Hilbert space consists of the empty-dot state $|0\rangle$ and the four single-electron eigenstates $|i\rangle$ with energies $E_i$ 
($i=1,\bar2,\bar3,4$). The probability vector describing the system is \cite{Brouwer,Timm,BM_Grifoni},
\begin{equation}
\mathbf{P}(t) =
(P_0(t),P_1(t),P_{\bar2}(t),P_{\bar3}(t),P_4(t))^{T},
\qquad
\sum_{i=0,1,\bar2,\bar3,4} P_i(t)=1 .
\end{equation}
\begin{figure}[thpb]
\centering
\includegraphics[width=0.7\textwidth]{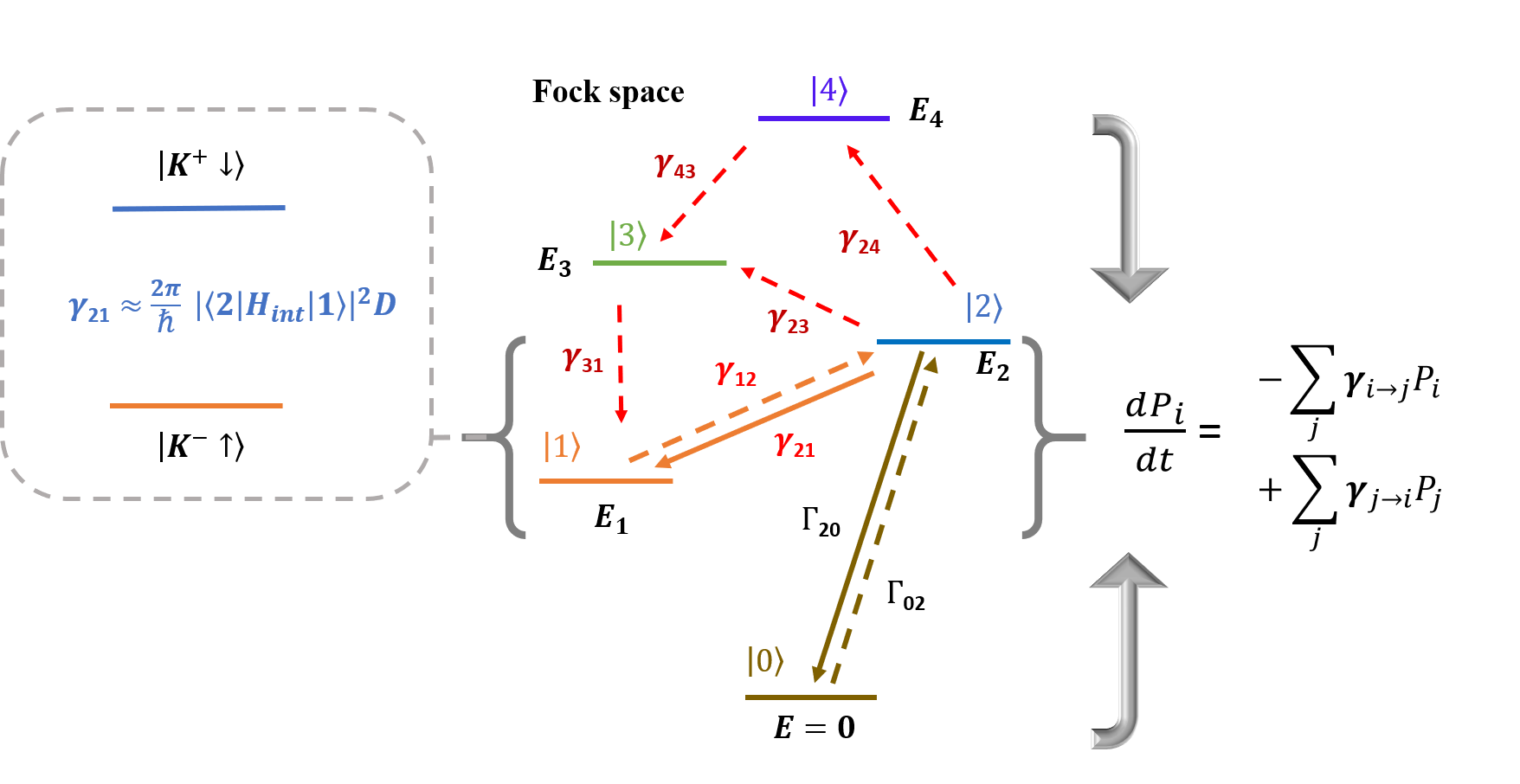}
\caption{Multilevel transition network used in the ME model. The BLG QD is described by the empty state $|0\rangle$ and four single-electron states $|1\rangle$--$|4\rangle$. Intrinsic relaxation processes between dot levels are characterized by rates $\gamma_{ij}$, while tunneling between the reservoir and the dot occurs with rates $\Gamma_{0i}$ and $\Gamma_{i0}$. The populations evolve according to the Pauli ME. Near spectral anticrossings the eigenstates hybridize, leading to modified transition rates that enter the rate matrix.}
\label{fig:Fig2_sup}
\end{figure}
\indent
During the load/wait stage of the experimental cycle the tunnel barrier to the reservoir is closed, so that no tunneling processes occur between the dot and the contact. 
The electron number in the dot therefore remains fixed, and the dynamics are governed entirely by intrinsic relaxation processes within the QD. 
These processes originate from the phonon and Johnson-noise mechanisms derived in Sec.~I.

Downhill transitions between dot levels are denoted by $\gamma_{ij}$ for processes $i\rightarrow j$ when $E_i>E_j$. 
The corresponding uphill transitions arise from thermally activated phonon re-excitation and obey detailed balance,
\begin{equation*}
\gamma_{ji}=\gamma_{ij}\,e^{-(E_i-E_j)/k_B T_e},
\qquad (E_i>E_j),
\end{equation*}
where $T_e$ is the electron temperature.

Because tunneling to the reservoir is suppressed in this regime, the empty-dot state remains unoccupied and
\begin{equation*}
P_0(t)\equiv0 .
\end{equation*}
\indent
The dynamics therefore reduce to the four-state single-electron subspace
\begin{equation*}
\mathbf{P}_{dot}(t) =
(P_1(t),P_{\bar2}(t),P_{\bar3}(t),P_4(t))^{T}.
\end{equation*}
\indent
The evolution of the populations is governed by the Pauli master equation
\begin{equation}
\dot{\mathbf{P}}_{dot}(t)
=
W_{dot}\mathbf{P}_{dot}(t),
\end{equation}
where $W_{dot}$ is the intra-dot transition matrix constructed from the intrinsic relaxation rates,
\begin{equation}
W_{dot}=
\begin{pmatrix}
-(\gamma_{12}+\gamma_{13}+\gamma_{14}) & \gamma_{21} & \gamma_{31} & \gamma_{41} \\
\gamma_{12} & -(\gamma_{21}+\gamma_{23}+\gamma_{24}) & \gamma_{32} & \gamma_{42} \\
\gamma_{13} & \gamma_{23} & -(\gamma_{31}+\gamma_{32}+\gamma_{34}) & \gamma_{43} \\
\gamma_{14} & \gamma_{24} & \gamma_{34} & -(\gamma_{41}+\gamma_{42}+\gamma_{43})
\end{pmatrix}.
\end{equation}
\indent
The structure of $W_{dot}$ ensures probability conservation, since the columns sum to zero. 
Consequently, the matrix possesses one eigenvalue equal to zero, corresponding to the stationary state. 
The remaining three eigenvalues $\{\lambda_1,\lambda_2,\lambda_3\}$ describe the relaxation modes of the coupled four-level system.\\
The general solution of the master equation can therefore be written as
\begin{equation}
P_i(t)
=
P_i^{(\infty)}
+
\sum_{j=1}^{4} c_{ij} e^{-\lambda_j t},
\qquad i=1,\bar2,\bar3,4 ,
\end{equation}
where $P_i^{(\infty)}$ are the stationary probabilities. 
Since the system relaxes thermally within the dot subspace, the stationary distribution is given by the Boltzmann weights
\begin{equation*}
P_i^{(\infty)}=
\frac{e^{-E_i/k_B T_e}}
{\sum_{m=1}^{4}e^{-E_m/k_B T_e}},
\qquad
P_0^{(\infty)}=0 .
\end{equation*}
\indent
The coefficients $c_{ij}$ are determined by the initial condition set during the load stage. 
For example, if the electron is initialized in state $\bar2$, the initial populations are
\begin{equation*}
P_1(0)=0,\quad
P_{\bar2}(0)=1,\quad
P_{\bar3}(0)=0,\quad
P_4(0)=0 .
\end{equation*}
\indent
\indent
If phonon re-excitation processes were neglected, the decay of the ES $\bar{2}$ would be governed by the bare escape rate,
\begin{equation*}
\gamma_{\bar{2}}^{\mathrm{bare}}
=
\gamma_{21}+\gamma_{23}+\gamma_{24}.
\end{equation*}
This expression represents the total intrinsic transition probability out of state $\bar{2}$ in the absence of repopulation from other levels.

However, near the spectral anticrossing the ES $\bar{2}$ and $\bar{3}$ become closely spaced in energy. As $B_\perp$ approaches the anticrossing point, the energy separation $E_{\bar3}-E_{\bar2}$ can become comparable to the thermal energy $k_B T_e$. In this regime thermally activated transitions between the two ESs become significant, allowing population to be transferred from $\bar{2}$ to $\bar{3}$ and vice versa.\\
\indent Consequently, the decay dynamics can no longer be described as the relaxation of a single isolated level. Instead, the ES population evolves through the coupled dynamics of the nearly degenerate states $\bar{2}$ and $\bar{3}$, together with their relaxation pathways to lower-energy states. The time evolution of the ES occupation therefore becomes multi-exponential,
\begin{equation*}
P_{\bar{2}}(t)
=
P_{\bar{2}}^{(\infty)}
+
\sum_{j=1}^{3}
A_j e^{-\lambda_j t},
\end{equation*}
where $\lambda_j$ are the non-zero eigenvalues of the full rate matrix describing the coupled four-level system.\\
\indent In this regime the experimentally observed decay rate is determined by the slowest relaxation mode of the rate matrix rather than by the escape rate of any individual transition channel. The effective relaxation rate is therefore
\begin{equation*}
\gamma_{\mathrm{effective}}(B_\perp)
=
\min\{\lambda_1,\lambda_2,\lambda_3\},
\end{equation*}
which corresponds to the smallest non-zero eigenvalue of the intra-dot transition matrix.\\
\indent Importantly, $\gamma_{\mathrm{effective}}$ is not associated with relaxation between any particular pair of states and cannot be obtained from a simple combination of the microscopic transition rates $\gamma_{ij}$. Instead it emerges from the collective dynamics of the coupled multilevel system, where population can repeatedly redistribute between the nearly degenerate ES before relaxing to the GS.

Near the state anticrossings, the proximity of higher excited states enables thermally activated population exchange, causing the observed lifetime to reflect the slowest relaxation mode of the coupled multilevel system. Consequently, although the qubit transitions remain spectroscopically well resolved, the lifetime extracted from the measurements ($T_{1m}$) at these operating points corresponds to this effective decay rate rather than to the intrinsic $T_1$ lifetime. 

In the numerical simulations presented in the main text, the $B_\perp$-dependent intrinsic rates $\gamma_{ij}$ derived in Secs.~I--III are used to construct the rate matrix $W_{\mathrm{dot}}(B_\perp)$. Solving the ME yields the population dynamics and the corresponding $\gamma_{\mathrm{eff}}$. Near the anticrossing, the proximity of ES modifies the decay dynamics: rather than a single exponential relaxation to zero population, the excited-state occupation approaches its Boltzmann value on a timescale determined by the smallest non-zero eigenvalue of the rate matrix.
\end{document}